\documentclass{article}

\usepackage[english]{babel}

\usepackage[letterpaper,top=2cm,bottom=2cm,left=2.5cm,right=2.5cm,marginparwidth=1.75cm]{geometry}

\usepackage{footmisc}
\usepackage{appendix}
\usepackage{amsmath}

\usepackage{float}
\usepackage{graphbox}
\usepackage{authblk}
\usepackage[colorlinks=true, allcolors=blue]{hyperref}
\usepackage{subcaption}
\usepackage{indentfirst}
\usepackage{graphicx}
\usepackage{comment}
\usepackage{multirow}
\usepackage{xcolor}
\usepackage[figurename=Fig.]{caption}
\usepackage{cite}
\usepackage{amsfonts}

\title{Quadratic speed-ups in quantum kernelized binary classification}
\author[1]{\normalsize Jungyun Lee}
\author[2,3,*]{\normalsize Daniel K. Park}
\affil[1]{\small \textit{Department of Physics, Yonsei University, Wonju, Republic of Korea}}
\affil[2]{\small \textit{Department of Applied Statistics, Yonsei University, Seoul, Republic of Korea}}
\affil[3]{\small \textit{Department of Statistics and Data Science, Yonsei University, Seoul, Republic of Korea}}

\date{}

\begin{document}
\maketitle
\footnotetext{\small * : corresponding author}
\footnotetext{\small Email addresses : dkd.park@yonsei.ac.kr}
\vspace{-5mm}
\begin{abstract}
Classification is at the core of data-driven prediction and decision-making, representing a fundamental task in supervised machine learning. Recently, several quantum machine learning algorithms that use quantum kernels as a measure of similarities between data have emerged to perform binary classification on datasets encoded as quantum states. The potential advantages of quantum kernels arise from the ability of quantum computers to construct kernels that are more effective than their classical counterparts in capturing patterns in data or computing kernels more efficiently. However, existing quantum kernel-based classification algorithms do not harness the capability of having data samples in quantum superposition for additional enhancements. In this work, we demonstrate how such capability can be leveraged in quantum kernelized binary classifiers (QKCs) through Quantum Amplitude Estimation (QAE) for quadratic speed-up. Additionally, we propose new quantum circuits for the QKCs in which the number of qubits is reduced by one, and the circuit depth is reduced linearly with respect to the number of sample data. We verify the quadratic speed-up over previous methods through numerical simulations on the Iris dataset.
\end{abstract}

\section{Introduction}

In the contemporary age of big data, machine learning (ML) has become an integral part of modern technology as an effective tool for making predictions and decisions based on data. Challenges arise when addressing problems involving complex, high-dimensional, and large datasets, which typically require advanced learning algorithms and significant computational power. Quantum information processing (QIP) holds promise for advancing ML beyond the limitations of classical methods. The advantages of quantum algorithms in solving certain computational tasks~\cite{HHL,rebentrost2014quantum,lloyd2014quantum,doi:10.1098/rspa.2015.0301}, the ability of quantum computers to efficiently harness exponentially large quantum state spaces~\cite{schuld2019quantum,havlivcek2019supervised,liu2021rigorous}, and the potential speed-ups in generating certain probability distributions~\cite{arute2019quantum,cerezo2020variational}, have catalyzed the development of quantum machine learning (QML)~\cite{wittek2014quantum, schuld2015introduction, biamonte2017quantum, zhang2020recent}.

One of the simplest and most well-established approaches in QML is based on the kernel method. This is because mapping classical data into the quantum Hilbert space naturally enables the computation of a kernel function on a quantum computer. This approach is often referred to as the quantum kernel method (QKM)~\cite{havlivcek2019supervised, mengoni2019kernel, schuld2021quantum}. In both classical and quantum kernel methods for ML, the central operation is measuring the similarity between two data points through a kernel function to extract and characterize patterns of the data for effective classification or prediction~\cite{hofmann2008kernel,schuld2019quantum}. The computational advantages of the QKM stem from the state and measurement postulates of quantum mechanics, which allow for efficient computation of certain kernel functions on a quantum computer. In particular, the Hadamard or swap test can be employed to exponentially speed-up the computation of fidelity between two quantum states, each representing data points to be compared~\cite{buhrman2001quantum, aharonov2006polynomial}. The Hadamard and swap tests have been harnessed in several quantum kernelized classification algorithms for exponential speed-ups with respect to the number of features in the data when evaluating the classification score~\cite{schuld2017implementing, blank2020quantum, PARK2020126422, Blank_2022, Pillay2024, DEOLIVEIRA2024127356}. These algorithms are known as the Hadamard classifier (HC) and swap test classifier (SC), respectively, and represent one of the simplest QML protocols with quantum speed-up. However, these algorithms do not provide any quantum advantage concerning the number of data samples, even though they require preparing an input quantum state containing all samples in quantum superposition~\cite{PARK2020126422}. Although the ability to encode the full dataset in superposition is a distinct feature in quantum computations, in previous works, it merely increased the size of the quantum circuits without yielding any computational advantages.

In this work, we present a protocol for integrating Quantum Amplitude Estimation (QAE)~\cite{brassard2002quantum} into quantum kernelized binary classifiers (QKCs), which are composed of HC and SC, resulting in a quadratic speed-up with respect to the number of data samples used in superposition. Additionally, we introduce simplified versions for both HC and SC by modifying the encoding process and the measurement scheme. This modification results in a reduction of one qubit and linearly reduces the circuit depth with respect to the number of sample data. Moreover, our proposed method does not rely on classical data processing, such as standardization or post-selection schemes, as required in the seminal work by Schuld et al.~\cite{schuld2017implementing}. We note in passing that while this work primarily focuses on the quantum speed-up with respect to the number of data samples in computing the classification score in QKCs, QAE can also be employed to expedite the estimation of kernel matrix elements. The latter is particularly beneficial in quantum-classical hybrid ML, where the quantum kernel matrix is utilized in classical ML algorithms, such as the support vector machine (SVM)~\cite{SVM1995, havlivcek2019supervised, huang_power_2021}.

The remainder of this paper is organized as follows. Sec.~\ref{sec:QKCs} describes the QKCs, the main focus of this work, and provides a brief review on HC and SC developed in Refs.~\cite{schuld2017implementing,blank2020quantum} to construct QKCs on a quantum computer. Sec.~\ref{sec:classification without superposition} explains how previous QKCs fail to utilize the ability to encode the entire dataset in superposition, thereby motivating the development of this work. Sec.~\ref{sec:QAE} provides a quick overview of QAE, a key ingredient of the main protocol presented in this work. In Sec.~\ref{sec:result}, first we present generalized and simplified QKCs developed in this work in Sec.~\ref{sec:modified classifiers}. Then describes how to integrate QAE into these methods for achieving the quadratic speed-up compared to the previous QKCs in Sec.~\ref{sec:method}. The simulation results to verify the improvements achieved by our method are detailed in Sec.~\ref{sec:simulation}. Additionally, we discuss the application of the maximum likelihood Quantum Amplitude Estimation (MLQAE)~\cite{suzuki2020amplitude} to our classifiers, and the maximum likelihood estimation (MLE) method for post-processing QAE results~\cite{grinko2021iterative}. The conclusion and outlook of our work are discussed in Sec.~\ref{sec:conclusion}.

\section{Quantum kernelized binary classifiers}
\label{sec:QKCs}

The primary objective of supervised binary classification is to accurately predict the label of the test data $\tilde{y}$ based on the labeled sample (training) dataset $\mathcal{D} = \lbrace(x_0,y_0),...,(x_{M-1},y_{M-1})\rbrace\in\mathbb{C}^N\times\lbrace 0,1\rbrace$, where $N$ is the number of features. A kernelized binary classifier solves this problem by utilizing the classification score, which is computed as
\begin{equation}
\label{eq:classificationscore}
  f(\tilde{x}) = \sum_{m=0}^{M-1}(-1)^{y_m} w_m k(x_m,\tilde{x}),
\end{equation}
where $k(x_m,\tilde{x})$ is the kernel function that quantifies the similarity between a sample data point $x_m$ and the test data $\tilde{x}$, $y_m\in\lbrace 0,1\rbrace$ is the label for $x_m$, and $w_m\in\mathbb{R}$ denotes the nonnegative weights assigned to the training samples. Subsequently, the label of the test data is predicted as $\tilde{y} = \mathrm{sign}(f(\tilde{x}))$. On QKCs, the classification score can be computed by applying either the Hadamard test~\cite{aharonov2006polynomial} or the swap test~\cite{buhrman2001quantum} on a quantum state that is prepared in a specific format. These methods yield classifiers known as the Hadamard classifier (HC)~\cite{schuld2017implementing} and swap test classifier (SC)~\cite{blank2020quantum}, respectively. In these scenarios, the kernel function is derived from the fidelity of two quantum states, each representing the training data and the test data. The potential advantage of such QKCs, meaning both HC and SC, lies in the effectiveness of the quantum kernel in capturing patterns in data and the ability of quantum computers to compute it efficiently compared to classical counterparts~\cite{schuld2017implementing,havlivcek2019supervised,blank2020quantum,PARK2020126422,liu2021rigorous,Blank_2022}.

QML algorithms, including QKCs, are naturally well-suited for datasets that are intrinsically quantum~\cite{10.1038/s43588-022-00311-3}, e.g. a final state of a quantum system prepared through certain quantum-mechanical processes. However, they can also be applied to classical data once the data is represented as a quantum state through the process known as quantum feature mapping~\cite{havlivcek2019supervised,schuld2019quantum,hur2023neural}. Without loss of generality, we use amplitude encoding~\cite{PhysRevA.102.032420,PhysRevA.101.032308,9259210,araujo_divide-and-conquer_2021, araujo_configurable_2023} as an example throughout the paper for encoding classical data into a quantum state. This representation expresses training and test data points, each consisting of $N$ features, as $|x_m\rangle:=\sum^{N}_{i=1}x_{m,i}|i\rangle$ and $|\tilde{x}\rangle:=\sum^{N}_{i=1}\tilde{x}_i|i\rangle$, respectively, using $O\left(\log_2(N)\right)$ qubits. However, we emphasize that QKCs are applicable to any datasets, as long as they are supplied as quantum states that can be handled by a quantum computer, either by an inherently quantum-mechanical system or through quantum feature mapping. 

\subsection{Hadamard classifier}
\label{sec:HC}
The HC was initially introduced in~\cite{schuld2017implementing} and garnered attention due to its relatively simple setup. More recent works, such as~\cite{blank2020quantum} and~\cite{Blank_2022}, have presented modifications to the HC in terms of a measurement scheme and encoding strategy, respectively. While we follow the description provided in~\cite{blank2020quantum} to explain the HC, we also elaborate on the differences from the original HC. The data set $\mathcal{D}$ is encoded in a quantum state as,
\begin{equation}
  |\Psi^h\rangle = \frac{1}{\sqrt{2}}\sum^{M-1}_{m=0}\sqrt{w_m}\Big[|0\rangle|x_m\rangle + |1\rangle|\tilde{x}\rangle\Big]|y_m\rangle|m\rangle,
\label{eq:HC state preparation}
\end{equation}
where the superscript $h$ indicates that the state pertains to HC, and the first register is an ancilla qubit, and the second resister $|x_m\rangle$ and $|\tilde{x}\rangle$ indicate training and test data, respectively, and $|y_m\rangle\in\{0,1\}$ represents the class of the $m$th training data. The original HC employs uniform weights, denoted as $w_m = 1/M \; \forall m\in[0,M-1]$, while more recent works have considered non-uniform weights for greater generality, subject to the constraint $\sum_m w_m = 1$. The last register denotes the index register, which flags each training data. This work can be achieved by using $n$ number of index qubits for $M=2^n$ number of training data. Besides, the class of the training data is represented via NOT gate controlled by index register. After preparing the state shown in Eq.~(\ref{eq:HC state preparation}), HC employ the Hadamard gate to the ancilla qubit $(H_a)$ to interfere the training data with the test data and construct the kernel to compare each similarity. It generate the quantum state $H_a|\Psi^h\rangle$, expressed as
\begin{equation}
  H_a|\Psi^h\rangle = \frac{1}{2}\sum^{M-1}_{m=0}\sqrt{w_m}\Big[|0\rangle|\psi^h_+(m)\rangle + |1\rangle|\psi_-^h(m)\rangle\Big]|y_m\rangle|m\rangle,
\end{equation}
where $|\psi^h_\pm(m)\rangle=|x_m\rangle\pm|\tilde{x}\rangle$. Finally, performing measurements on the ancilla and class qubit can realize kernel-based binary classification. Although the original HC used a conditional measurement selecting the branch with the ancilla in state $|0\rangle$~\cite{schuld2017implementing}, the classification result can be obtained by measuring the expectation value of a two-qubit observable, reducing the number of experiments by about a factor of two. This can be expressed as follows:
\begin{equation}
  \langle\sigma_z^a\sigma_z^c\rangle = \sum^{M-1}_{m=0}(-1)^{y_m}w_m\Re(\langle x_m|\tilde{x}\rangle),
  \label{eq:HC two measurement observable}
\end{equation}
where the superscript $a$ ($c$) indicates that the measurement operator acts on the ancilla (class) qubit and $\Re(\cdot)$ indicates the real part~\cite{blank2020quantum}. The expectation value assigns the class of the test data $\tilde{y}$ as 0 when the expectation result is positive and 1 when it is negative.

\subsection{Swap test classifier}
\label{sec:SC}
The SC first proposed in~\cite{blank2020quantum} functions similarly to HC. However, SC is based on measuring the full quantum state fidelity, whereas HC uses only the real part of the state overlap. In other words, the primary difference between SC and HC lies in the fact that the former fully utilizes the quantum feature map by considering both the real and imaginary parts. The majority of the state preparation processes are also similar to HC, however, unlike HC, the test data and training data are encoded in different registers. Thereby the ancilla qubit can be retreated from the preparation process, 
\begin{equation}
  |\Psi^s\rangle = \sum^{M-1}_{m=0}\sqrt{w_m}|0\rangle|x_m\rangle|\tilde{x}\rangle|y_m\rangle|m\rangle,
  \label{eq:SC state preparation}
\end{equation}
where the superscript $s$ indicates that the state pertains to SC, and the first register is an ancilla qubit, the second and the third registers denote the training data register and the test data register, respectively, and $|y_m\rangle\in\{0,1\}$ represents the class of $m$th training data and the last one indicates index register. After the encoding process, in contrast to HC, which applies the Hadamard gate, a sequence of gates, $H_a\cdot \text{c-swap}(a;x_m,\tilde{x})\cdot H_a$, is applied to $|\Psi^s\rangle$. Here, $H_a$ represents the application of the Hadamard gate to the ancilla, and the $\text{c-swap}(a;x_m,\tilde{x})$ is the application of the swap gate between the training data qubit and the test data qubit controlled by the ancilla qubit. The quantum state $|\Psi^s\rangle$ after this interference is 
\begin{equation}
  H_a\cdot \text{c-swap}(a;x_m,\tilde{x})\cdot H_a|\Psi^s\rangle = \frac{1}{2}\sum^{M-1}_{m=0}\sqrt{w_m}\Big[|0\rangle|\psi^s_+(m)\rangle + |1\rangle|\psi_-^s(m)\rangle\Big]|y_m\rangle|m\rangle,
\end{equation}
 where $|\psi^s_\pm(m)\rangle=|x_m\rangle|\tilde{x}\rangle\pm|\tilde{x}\rangle|x_m\rangle$. The SC is finalized with the two-qubit measurements, which results in
\begin{equation}
  \langle\sigma_z^a\sigma_z^c\rangle = \sum^{M-1}_{m=0}(-1)^{y_m}w_m\big|\langle x_m|\tilde{x}\rangle\big|^2,
  \label{eq:SC two measurement observable}
\end{equation}
where the superscript $a$ ($c$) indicates that the measurement operator acts on the ancilla (class) qubit. The sign of the expectation value determines the class of the test data.

\section{Classification without data superposition}
\label{sec:classification without superposition}

The ability to encode the entire dataset in a quantum superposition state is one of the unique properties of quantum computing~\cite{simon1997power,nielsen2000quantum}. In QKCs, placing $M$ training data in superposition is achieved by introducing the index register consisting of $\log_2 (M)$ qubits as shown in Eqs.~(\ref{eq:HC state preparation}) and (\ref{eq:SC state preparation}).
To implement this state preparation routine manually, it requires $M$ unitary data encoding gates applied to the data registers, controlled by $\log_2(M)$ index qubits. This setup ensures that each data sample is entangled with one computational basis state of the index register. In this scenario, if the number of training data increases by $K$, the process necessitates an additional $\log_2(K)$ index qubits. Consequently, it involves a total of $(M+K)$ unitary gates controlled by $\log_2(MK)$ qubits, resulting in a linear increase in circuit depth~\cite{da2022linear}. In summary, increasing the number of training data in the superposition state leads to a logarithmic increase in circuit width and a linear increase in circuit depth.

However, since all quantum operations are performed solely on the ancilla and class qubits once the initial states are provided, the presence of the index register and the preparation of data superposition do not contribute to computing Eqs. (\ref{eq:HC two measurement observable}) and (\ref{eq:SC two measurement observable}). To see this more clearly, let us express the initial state of QKCs as
\begin{equation}
    |\Psi^l\rangle = \sum^{M-1}_{m=0}\sqrt{w_m}|\psi^l(m)\rangle|y_m\rangle|m\rangle,
\end{equation}
where $l\in\lbrace h,s\rbrace$ is the label for indicating whether the state is for HC or SC, and $|\psi^h(m)\rangle = (|0\rangle|x_m\rangle + |1\rangle|\tilde{x}\rangle)/\sqrt{2}$ and $|\psi^s(m)\rangle = |0\rangle|x_m\rangle|\tilde{x}\rangle$ represent the quantum state that contains the ancilla, the $m$th training data and the test data. 
Since all subsequent quantum operations are applied only to the ancilla qubit and data register, which are contained in $|\psi^l(x_m,\tilde{x})\rangle$, and the class qubit, one can trace out the index register as it is completely ignored in all subsequent steps. The partial trace of the index register yields the mixed state
\begin{equation}
    \rho(x_m,\tilde{x},y_m) = \sum_{m=0}^{M-1}w_m |\psi^l(m)\rangle\langle\psi^l(m)|\otimes |y_m\rangle\langle y_m|.
    \label{eq:mixed state without index}
\end{equation}
This means that one cannot differentiate between the procedure outlined in previous sections and an alternative approach that performs Hadamard or swap test on the mixed state $\rho(x_m,\tilde{x},y_m)$, even though they are physically different. 
Furthermore, the latter process is equivalent to a protocol that independently performs the weighted Hadamard or swap test on the state $|\psi^l(m)\rangle$ for each sample data, and aggregates the total of $M$ outcomes classically (see~\ref{sec:weighted Hadamard and swap test}). This procedure is more practical than the original HC and SC, as each circuit does not require the index register consisting of $\log_2 (M)$ qubits and $M$ controlled-gates, controlled by those index qubits, for encoding the entire dataset. Therefore, unless the dataset is provided in the form of $|\Psi^h\rangle$ in Eq.~(\ref{eq:HC state preparation}) or $|\Psi^s\rangle$ in Eq.~(\ref{eq:SC state preparation}), superposing the entire dataset makes the algorithm unnecessarily more complex without offering any computational advantage.

\section{Quantum Amplitude Estimation}
\label{sec:QAE}
This section briefly reviews Quantum Amplitude Estimation (QAE)~\cite{brassard2002quantum}, a key ingredient of the main protocol presented in this work. Since there are many comprehensive reviews on this topic~\cite{montanaro2015quantum,suzuki2020amplitude,grinko2021iterative,nakaji2020faster,intallura2023survey}, our intention is to provide a minimalistic overview necessary for understanding our protocol. 

QAE is the task of estimating the value of $a$ for the quantum state $|\psi\rangle = \sqrt{a}|s\rangle + \sqrt{1-a}|s_\perp\rangle$, where $a\in [0,1]$. This problem can be resolved by using two fundamental quantum algorithms that constitute QAE, namely Quantum Amplitude Amplification (QAA)~\cite{brassard2002quantum} and Quantum Phase Estimation (QPE)~\cite{nielsen2000quantum}.

\begin{itemize}
  \item QAA is a generalization of the Grover's search algorithm~\cite{grover1996fast} which finds the target state quadratically faster than classical counterparts. QAA amplifies the probability of finding a specific state in a quantum system even if they are not in uniform superposition like in Grover's algorithm.
  \item QPE is a quantum algorithm to estimate $\theta$, where $U|\psi\rangle = e^{2\pi i \theta}|\psi\rangle$, for a target unitary operator $U$ and its eigenvector $|\psi\rangle$. The precision of the estimation depends on the number of ancilla qubits ($t$) used, which is known as $t$-bit approximation.
\end{itemize}

\begin{figure}
  \centering
  \includegraphics[width=0.6\textwidth]{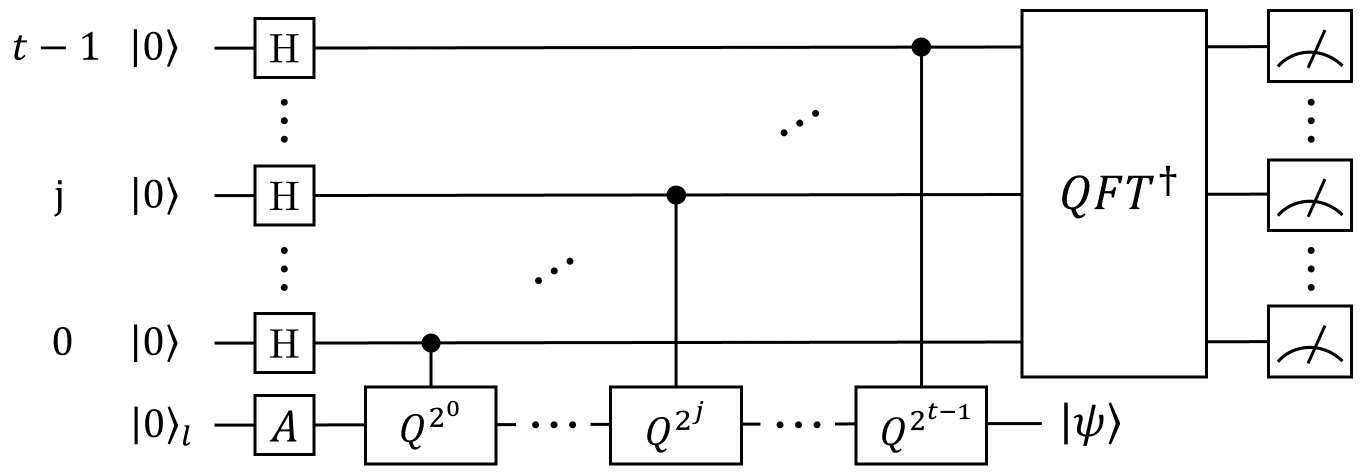}
  \caption{A quantum circuit diagram illustrating the implementation of the QAE algorithm, where $A$ represents the state preparation task, $Q$ denotes the Grover operator, and $QFT$ stands for Quantum Fourier Transform~\cite{nielsen2000quantum}.}
  \label{fig:qae-circuit}
\end{figure}

In simple terms, QAE can be viewed as a variant of QPE that uses the Grover operator ($Q$) as the target unitary operator. Therefore, the precision of the estimation can be adjusted by controlling the number of ancilla qubits, just as QPE. The mathematical procedure is described as follows:\\
When the state preparation operator is $A$, an initial state $|\psi\rangle$ is defined as
\begin{equation}
    A|0\rangle^{\otimes l} \equiv A|0\rangle_l \equiv |\psi\rangle = \sqrt{a}|s\rangle + \sqrt{1-a}|s_\perp\rangle = \sin(\theta)|s\rangle + \cos(\theta)|s_\perp\rangle,
  \label{eq:QAE initial state}
\end{equation}
where $0\leq\theta\leq\pi/2$. 
Since the main goal of QAE is to estimate $a$, $|s\rangle$ is called the ``good state" or ``target state". The Grover operator ($Q$), consists of the phase oracle ($U_f$), and diffusion operator ($V$):
\begin{equation}
  Q = VU_f \text{ where } U_f = I - 2|s\rangle\langle s| \text{ and } V = I - 2|\psi\rangle\langle\psi| = A\big(I - 2|0\rangle_l\langle0|_l\big)A^\dagger.
\end{equation}
The phase oracle, $U_f$, multiplies the good states by $-1$ and the whole state is reflected around the mean by $V$, which performs the amplification of the specific amplitude. When $|\psi_\pm\rangle$ is the eigenvector of $Q$ corresponding to eigenvalues $\lambda_\pm = e^{\pm2i\theta}$, the initial state $|\psi\rangle$ can be expressed with the eigenvectors of $Q$ as
\begin{equation}
  |\psi\rangle = \frac{-i}{\sqrt{2}}\Big(e^{2i\theta}|\psi_+\rangle+ e^{-2i\theta}|\psi_-\rangle\Big).
\end{equation}
Therefore, we can estimate $\theta$ or $-\theta$ using the QPE algorithm. Consequently, QAE can perform $t$-bit approximation of $a$, which is given by $\tilde{a} = \sin^2(\tilde{\theta}) = \sin^2\left(\pi y/2^t\right)$, through the quantum circuit depicted in Fig.~\ref{fig:qae-circuit}, where $y$ is non-negative and the decimal form of the most frequent measured state. For example, when the number of ancilla $t=2$, and if $|01\rangle$ is the most measured state, then $y$ is 1 and the $\tilde{a}=0.5$.

The estimator $\tilde{a}$ satisfies the following bounds where $a=\sin^2(\theta)$ and $\tilde{a}=\sin^2(\tilde{\theta})$,
\begin{equation}
  |a-\tilde{a}| \leq \frac{2\sqrt{a(1-a)}\pi}{N_q} + \frac{\pi^2}{N_q^2} \leq \frac{\pi}{N_q} + \frac{\pi^2}{N_q^2} = \mathcal{O}(N_q^{-1})
  \label{eq:qae error bound}
\end{equation}
with a probability of at least $8/\pi^2\;(\approx 81\%)$~\cite{brassard2002quantum}, where $N_q$ is the number of applications of $Q$, equal to $2^t$. Since $\tilde{a}$ is a probability, QAE is one of the point estimations of the specific probability. In classical point estimation, the error bound scales as $\mathcal{O} (1/\sqrt{N_c})$ for the number of classical sample $N_c$, whereas QAE achieves a scaling of $\mathcal{O}(1/N_q)$, which implies that QAE has the potential of quadratic speed-up over classical simulation.

\section{Results}
\label{sec:result}

In Sec.~\ref{sec:classification without superposition}, we demonstrated that QKCs failed to utilize the capability of placing training data in quantum superposition for computational advantage. In this section, we present new protocol for QKCs using QAE to leverage the ability of the QML algorithm to process the entire dataset in superposition. Before delving into the main protocol, we first show that QKCs can be simplified to reduce the size of quantum circuits by using a specific encoding process. Subsequently, we modify the measurement scheme to enhance their suitability for QAE. The main protocols that combine the simplified HC (SHC) and simplified SC (SSC) with QAE are presented in Sec.~\ref{sec:method}.

\subsection{Simplified quantum kernelized binary classifier}
\label{sec:modified classifiers}
In the original QKCs, the encoding strategy is that label the class of the arbitrary encoded training data into class qubit $|y_m\rangle$ by using NOT gates controlled by the index register $|m\rangle$~\cite{schuld2017implementing, blank2020quantum, Blank_2022}. Therefore, the number of multi-controlled gates for labeling the class of the training data increases linearly with respect to the number of class 1 data. For the simplification of this process we use an ordered encoding strategy which indicates encoding the training data in a specific order, e.g. encode class 1 data after class 0 data are encoded or encode class 0 data and class 1 data alternately. With this strategy, the class information of training data points is implicitly encoded into one of the index qubits. For brevity, we refer to this qubit as a class-identifiable qubit. Thus, labeling the class of the data points can be achieved by a CNOT gate from the class-identifiable qubit to the class qubit without increasing the number of control qubits or gates by data size. However, in fact, this CNOT gate just duplicates the information that is already in the class-identifiable qubit so that we can remove the class qubit and the work of it can be shifted to the class-identifiable qubit. Then measuring the ancilla and class-identifiable qubits provides the same work using one less qubit and linear reduction in circuit depth with respect to the number of training data. Furthermore, the measurement can be further reduced to a single-qubit measurement by the Clifford transformation. Without loss of generality, we encode the data with a class-order in the subsequent description, i.e. encode class 1 data after class 0 data are encoded, and call it as a class-ordered encoding, which can be performed by a unitary gate, $U_{co} (x_m,y_m)$. Furthermore, our description throughout the paper is focused on the balanced data in which the number of training data points in two classes is equal. However, this strategy is also applicable in the case of unbalanced data as long as $\sum_m w_m = 1$ since we don't need to use all the basis of the index register.
\begin{figure}[t]
\centering\begin{subfigure}{\textwidth}
 \centering
 \includegraphics[scale=0.5]{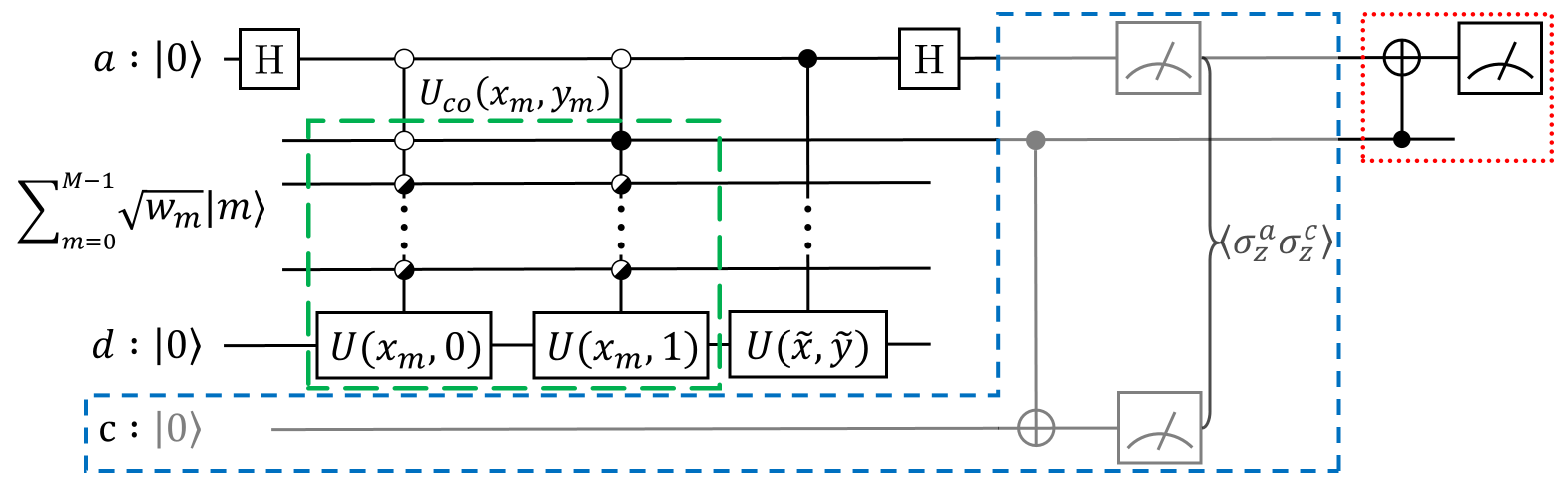}
 \caption{Simplified Hadamard classifier}
 \label{fig:SHC}
\end{subfigure}
\begin{subfigure}{\textwidth}
 \centering
 \includegraphics[scale=0.5]{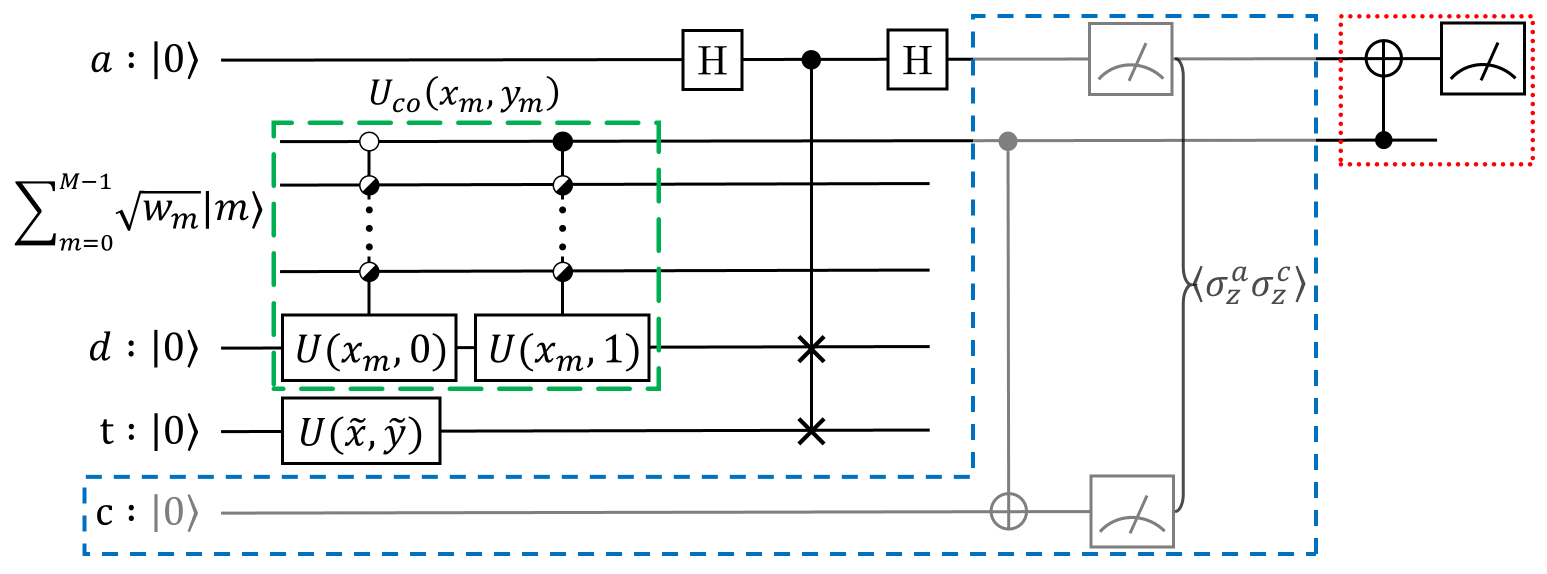}
 \caption{Simplified swap test classfier}
 \label{fig:SSC}
 \end{subfigure}
 \caption{Quantum circuit diagrams for (a) Simplified Hadamard classifier (SHC) and (b) Simplified swap test classifier (SSC). The first register is the ancilla qubit ($a$), and the second is the weighted index register. $M=2^n$ is an acceptable number of data for $n$ number of index qubits and $|m\rangle = |m_0 m_1 ... m_{n-1}\rangle$, where $m_k \in \{0,1\}$ $ \forall k=0,1,...,n-1$. There is a difference between SHC and SSC to encode the data. In SHC, the third register is the data qubit ($d$)  for encoding both training and test data. On the other hand, in SSC, the training data and test data are encoded in different registers, third register and fourth register in (b), namely the training data qubit ($d$) and test data qubit ($t$), respectively. The final register is the class qubit ($c$), which can be removed using our proposed encoding scheme, green long-dashed box.}
 \label{fig:SQKCs circuit}
\end{figure}
The Simplified QKCs (SQKCs) shown in Fig.~\ref{fig:SQKCs circuit} interfere the training data and test data which are initialized by $U(x_m, y_m)|0\rangle = |x_m\rangle \;\forall m=0,1,..., 2^n-1$, and $U(\tilde{x}, \tilde{y})|0\rangle = |\tilde{x}\rangle$, and predict the class of the test data $\tilde{y}$ by harnessing its superposition. The green long-dashed box, $U_{co}(x_m,y_m)$, indicates the class-ordered encoding endowing the ability to identify the class of the training data to the top qubit of the index register, $|m_0\rangle$, meaning that $|m_0\rangle$ becomes the class-identifiable qubit. The half-filled circles indicate that the unitary operation is the uniformly controlled gate~\cite{mottonen2004transformation, bergholm2005quantum, Blank_2022}. 
Therefore, the data set is prepared as,
\begin{equation}
 |\Phi^l\rangle = \sum^{M-1}_{m=0}\sqrt{w_m}|\psi^l(m)\rangle|m\rangle,
 \label{eq:simplified classifier state preparation state}
\end{equation}
where $l\in\lbrace h,s\rbrace$ is the label for indicating whether the state is for Simplified Hadamard classifier (SHC) or Simplified swap test classifier (SSC), and $|m\rangle = |m_0 m_1...m_{n-1}\rangle = |y_m m_1 m_2...m_{n-1}\rangle$.
The state after interference between training data and test data through the Hadamard gate or swap test is 
\begin{equation}
    \frac{1}{2} \sum_{m=0}^{M-1} \sqrt{w_m} \big[|0\rangle|\psi_+^l(m)\rangle+|1\rangle|\psi_-^l(m)\rangle \big]|m\rangle.
\label{eq:sQKCs after interference}
\end{equation}
The measurement scheme introduced in~\cite{blank2020quantum} using class-ordered encoding is depicted in the blue dashed box, which presents performing the classification using the expectation value of two-qubit observable. The red dotted box is our proposed measurement scheme removing class qubit and making the classification be realized with a single qubit. To be more precise, the class qubit can be discarded when shifting the $Z$-measurement of the class qubit to the class-identifiable qubit $|m_0\rangle$, and we can reduce the measurement operator by harnessing the NOT gate controlled by $|m_0\rangle$ where the target is ancilla qubit using the Clifford transformation, $I\otimes Z = (CNOT)(Z\otimes Z)(CNOT)^\dagger$~\cite{gottesman1998theory}. 
Since we use the class-ordered encoding, $m_0$ (hence, $y_m$) is 0 for $m \in [0,M/2-1]$, and 1 for $m\in[M/2,M-1]$. 
With this encoding, the final quantum state $|\Phi_f^l\rangle$ before measurement, which is obtained after the CNOT gate, is
\begin{align}
 |\Phi_f^l\rangle = \frac{1}{2}\Bigg{[}&|0\rangle \Bigg{\lbrace}
 \sum_{m=0}^{M/2-1}\sqrt{w_m}|\psi_+^l(m)\rangle|m\rangle +
 \sum_{m=M/2}^{M-1}\sqrt{w_m}|\psi_-^l(m)\rangle|m\rangle \Bigg{\rbrace}\nonumber\\ 
 + &|1\rangle \Bigg{\lbrace}
 \sum_{m=0}^{M/2-1}\sqrt{w_m}|\psi_-^l(m)\rangle|m\rangle +
 \sum_{m=M/2}^{M-1}\sqrt{w_m}|\psi_+^l(m)\rangle|m\rangle \Bigg{\rbrace}\Bigg{]},
 \label{eq:SQKC final state}
\end{align}
where the first qubit is the ancilla register and $|\psi_{\pm}^{l}(m)\rangle$ contains the training data point in class 0 when $m \in [0,M/2-1]$, and in class 1 when $m\in[M/2,M-1]$. The measurement result can be described with an expectation value of a one-qubit observable as, 
\begin{equation}
 \langle\sigma_z^{a}\rangle = \sum^{M-1}_{m=0}(-1)^{y_m}w_m k(x_m,\tilde{x}),
 \label{eq:SQKC one measurement observable}
\end{equation}
where $k(x_m, \tilde{x})$ is the kernel computed as $\Re\left( \langle x_m|\tilde{x}\rangle\right)$ for SHC and as $\big|\langle x_m|\tilde{x}\rangle\big|^2$ for SSC. This expectation value is equivalent to the one introduced in~\cite{blank2020quantum}. Thus, our SQKCs and the previous QKCs are themselves equivalent with the same classification score, Eq. (\ref{eq:SQKC one measurement observable}) with Eqs. (\ref{eq:HC two measurement observable}) and (\ref{eq:SC two measurement observable}), which assigns the class of the test data $\tilde{y}$ as 0 when the expectation result is positive and 1 when it is negative.

\subsection{Main protocols}
\label{sec:method}

Herein we elaborate on a method of applying QAE to SQKCs. Then, we introduce our comparison methodology between SQKCs with QAE (SQKCs-QAE) and those without it (SQKCs).

\paragraph{SQKCs-QAE algorithm}

The final quantum states of the SQKCs, SHC and SSC, can be expressed in the form of Eq. (\ref{eq:QAE initial state}) by rewritting Eq.~(\ref{eq:SQKC final state}) as
\begin{equation}
\label{eq:state_qae}
    |\Phi_f^l\rangle = \sqrt{\frac{1}{2}\left[1+\sum_{m=0}^{M-1}w_m(-1)^{y_m} k(x_m,\tilde{x})\right]}|0\rangle|\Phi_0\rangle + \sqrt{\frac{1}{2}\left[1-\sum_{m=0}^{M-1}w_m(-1)^{y_m} k(x_m,\tilde{x})\right]}|1\rangle|\Phi_1\rangle,
\end{equation}
where $|\Phi_0\rangle$ and $|\Phi_1\rangle$ are normalized states. It is evident from this expression that the classification score can be extracted from the probability of obtaining either 0 or 1 as the outcome of measuring the first qubit, which can be estimated through the use of QAE. For instance, if $\Pr(1) = \frac{1}{2}\left[1-\sum_{m}w_m(-1)^{y_m} k(x_m,\tilde{x})\right]$ is estimated via QAE, the decision rule for assigning the label to the test data is as follows: if the QAE result is smaller than 0.5, then set $\tilde{y}$ to 0; otherwise, set $\tilde{y}$ to 1. In this case, the phase oracle, $U_f$, of QAE is the single-qubit Pauli-$Z$ gate acting on the first qubit, since it multiplies the good state, which, in this case, is $|1\rangle|\Phi_1\rangle$, by $-1$.

\begin{figure}
 \centering
 \includegraphics[scale=0.5]{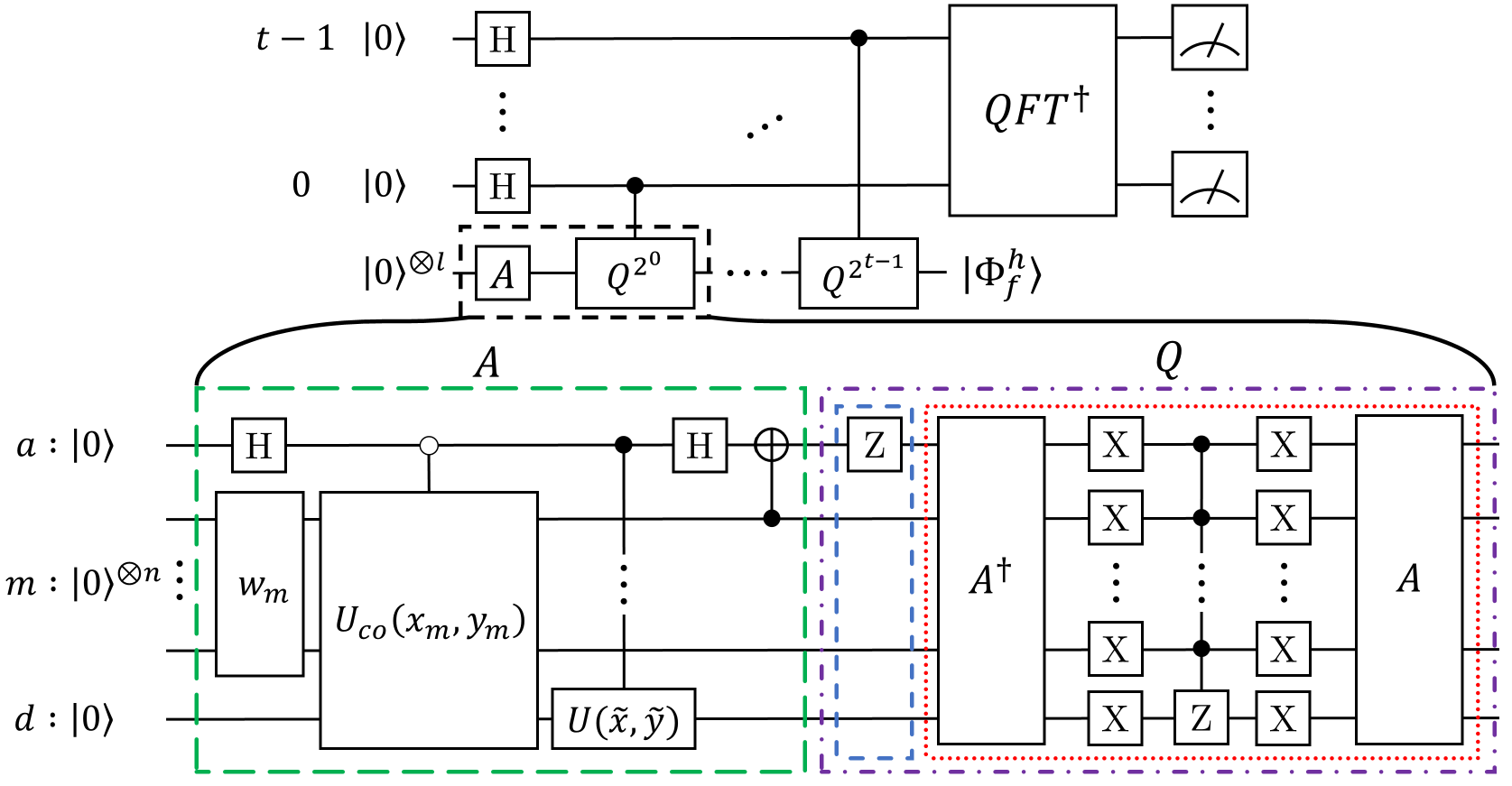}
\caption{Quantum circuit diagram for SHC-QAE, which estimates the probability amplitude of the final state of SHC with the ancilla in state $|1\rangle$ using QAE.
$A$ indicates the state preparation operator (green long-dashed box) and $Q$ indicates the Grover operator (purple dash-dotted box). The Grover operator consists of the phase oracle (blue dashed box) and diffusion operator (red dotted box).}
\label{fig:SHC-QAE}
\end{figure}

The quantum circuit of SHC-QAE, which combined SHC and QAE, is depicted in Fig.~\ref{fig:SHC-QAE}. To apply the QAE algorithm, we need to define the state preparation operator ($A$) and the Grover operator ($Q$). The $A$ operator corresponds to either SHC or SSC circuits, which prepares the final state as shown in Eq.~(\ref{eq:state_qae}). After the state preparation, the quantum state undergoes the phase oracle ($U_f$), which can be easily implemented by applying $Z$ gate to the ancilla, since its role is to multiply $-1$ to the target state $|1\rangle |\Phi_1\rangle$ in the case of estimating $\Pr(1)$. On the other hand, if we set the oracle with the sequence of $XZX$ gate, QAE can estimate the amplitude of $|0\rangle|\Phi_0\rangle$. The second component of the Grover operator, namely the diffusion operator ($V$), has a specific structure with a multi-controlled $Z$ gate sandwiched by $X$ gates just as shown in the red dotted box in Fig.~\ref{fig:SHC-QAE}. Using this implementation of $A$ and $Q$, the $t$-bit approximation of the amplitude for state $|1\rangle$ can be performed by means of the QPE structure.

\paragraph{Performance measure}

\begin{table}[h]
\setlength{\tabcolsep}{7pt}
\renewcommand{\arraystretch}{1.5}
\centering
\begin{tabular}{|c|c|c|}
\hline
& Number of samples & Error \\ \hline
SQKCs-QAE& $2^{t+1}N^{q}_{shot} := 2^{t+1}$& \multirow{2}{*}{\begin{tabular}[c]{@{}c@{}} The $81$st largest value among all errors, $|a-\tilde{a}_i|,\;i=1,2,...,I$.\\  $I$ is the number of repetitions for a given number of samples. \end{tabular}} \\ \cline{1-2}
SQKCs& $N^{c}_{shot} := 2^{t+1}$&\\ \hline
\end{tabular}
 \caption{An overview of variables used for comparing the performance of SQKCs-QAE and SQKCs.}
 \label{table:comparison method}
\end{table}

To validate the quadratic speed-up through numerical simulations, it is crucial to establish a clear and rigorous evaluation criterion for comparing the performance of SQKCs-QAE and SQKCs. In this regard, we compare the rate at which the estimation errors decrease in SQKCs-QAE and SQKCs as the number of samples increases. Two variables, namely the number of samples and the estimation error, whose relationship is analyzed and compared in the subsequent section, are summarized in Table~\ref{table:comparison method}.

Firstly, we define the term ``sample" to represent a query to the state preparation oracle, $A$. In other words, the number of samples corresponds to the instances of applying $A$ that generate the output state of SQKCs, which is given in Eq. (\ref{eq:SQKC final state}). In the case of SQKCs-QAE, there are $2^{t+1} - 2 \approx 2^{t+1}$ number of samples, when $t$ ancilla qubits are used for QAE. This is because there are $\sum^{t-1}_{x=0} 2^x = 2^t -1$ number of $Q$ for the entire process (see, Fig.~\ref{fig:qae-circuit}), and each Grover operator involves 2 instances of the state preparation, $A$ and $A^\dagger$ in $V = A\big(I - 2|0\rangle_m\langle0|_m\big)A^\dagger$, where $Q = VU_f$. For SQKCs, there are no additional operations like the Grover operator. Therefore, we can increase the number of samples by increasing the number of shots, $N^c_{shots}$. Certainly, this principle also extends to the SQKCs-QAE, as well. Thus, the number of samples, in SQKCs-QAE case, becomes $2^{t+1}N^q_{shot}$. Since this value must be equal for precise comparison, we set the number of shots for SQKCs-QAE, $N^q_{shots}$, to 1 and $N^c_{shots}$ to $2^{t+1}$. Consequently, the number of samples for both SQKCs-QAE and SQKCs becomes $2^{t+1}$.
Next, we define the error as $|a-\tilde{a}_i|$, where $a$ and $\tilde{a}_i$ denote the real value we want to estimate and its $i$th estimator, respectively. QAE satisfies the error bound, Eq. (\ref{eq:qae error bound}), with a probability of at least $8/\pi^2\;(\approx 81\%)$. This means, in other words, that if 1000 circuits are generated with each number of samples, and each is measured to produce QAE error results, thereby resulting in 1000 QAE error results, at least 810 of them will satisfy the bound. Note that, in this scenario, $I$ in Table~\ref{table:comparison method} is 1000.
Therefore, repeating SQKCs-QAE and SQKCs multiple times for each number of samples, creating multiple error results at each sample size, sorting them, then, comparing the 81\% maximum error of SQKCs-QAE and SQKCs at each sample constitutes a valid comparison. By repeatedly performing QAE and utilizing the method of estimating the 81st percentile, the success probability of QAE can be boosted to nearly 100\%. In conclusion, by comparing the error scaling of SQKCs-QAE and SQKCs we can verify whether QAE achieves any speed-up in SQKCs.
Moreover, for a more concise comparison, we fitted each error result to a linear curve using the $\log_2$ function (using the $\log_2$ function is justified since the number of samples increases as a power of 2). The slope of the fitted line for SQKCs-QAE and SQKCs indicate how fast the estimation error decreases. Thus, we can verify the speed-up quantitatively by investigating the ratio between two slopes, namely (slope of the linear fit for SQKCs-QAE estimation error)/(slope of the linear fit for SQKCs estimation error), which we refer to as the slope ratio.

\subsection{Numerical simulation}
\label{sec:simulation}
\subsubsection{SQKCs-QAE vs SQKCs}
We verify the quadratic speed-up achieved by our SQKCs-QAE over SQKCs through numerical simulations on the Iris dataset. For the simulation, we utilize the first and last classes, \textit{setosa} and \textit{virginica} (referred to as class 0 and class 1 in this paper), based on two features of the Iris dataset: \textit{sepal width} and \textit{petal length}. Specifically, we illustrate the performance of the classifiers using the following simple dataset, consisting of two training data points (one from class 0 and the other from class 1) with uniform weights $w_m = 1/2$, along with one test data point:
\begin{equation}
\begin{array}{cc}
 |x_0\rangle= 0.9635|0\rangle + 0.2676|1\rangle\text{, Iris sample 23 : class 0}\\
 \;\,|x_1\rangle= 0.3526|0\rangle + 0.9358|1\rangle\text{, Iris sample 119 : class 1}\\
 \;\;\;\;|\tilde{x}\rangle= 0.3856|0\rangle + 0.9227|1\rangle\text{, Iris sample 123 : class 1}.\\
\end{array}
\label{eq:QAE data}
\end{equation}
The SHC and SSC results for the data in Eq. (\ref{eq:QAE data}) are approximately Pr(1) = 0.5952 and 0.6541, respectively. This implies that the test data can be successfully classified as $\tilde{y} = 1$ through both SHC and SSC, provided that $\Pr(1)$ is estimated with high accuracy.

\begin{figure}[t]
\centering
\begin{subfigure}{0.45\textwidth} 
 \centering
 \includegraphics[scale=0.4]{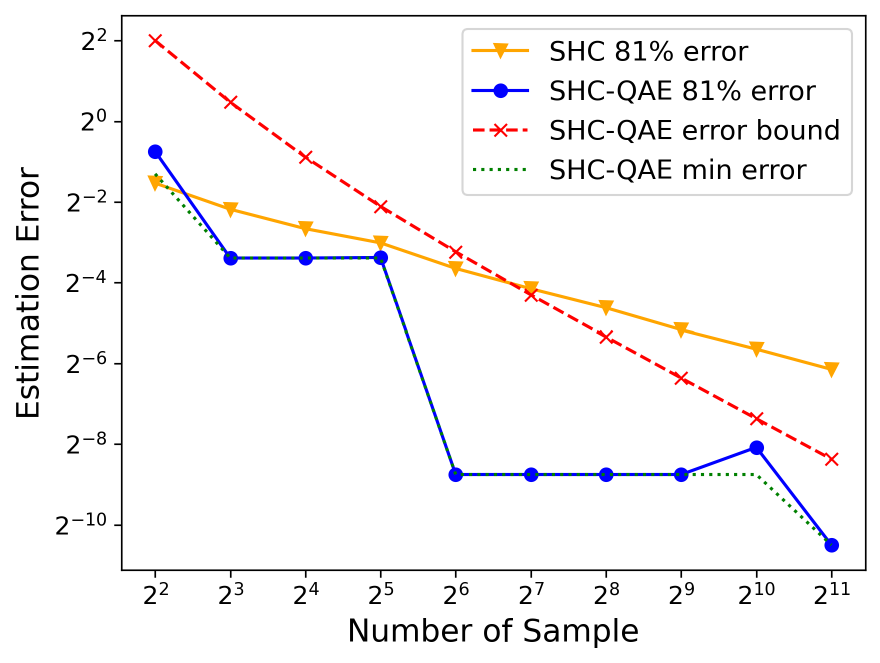}
 \caption{SHC comparison result}
\end{subfigure}
\begin{subfigure}{0.45\textwidth} 
 \centering
 \includegraphics[scale=0.4]{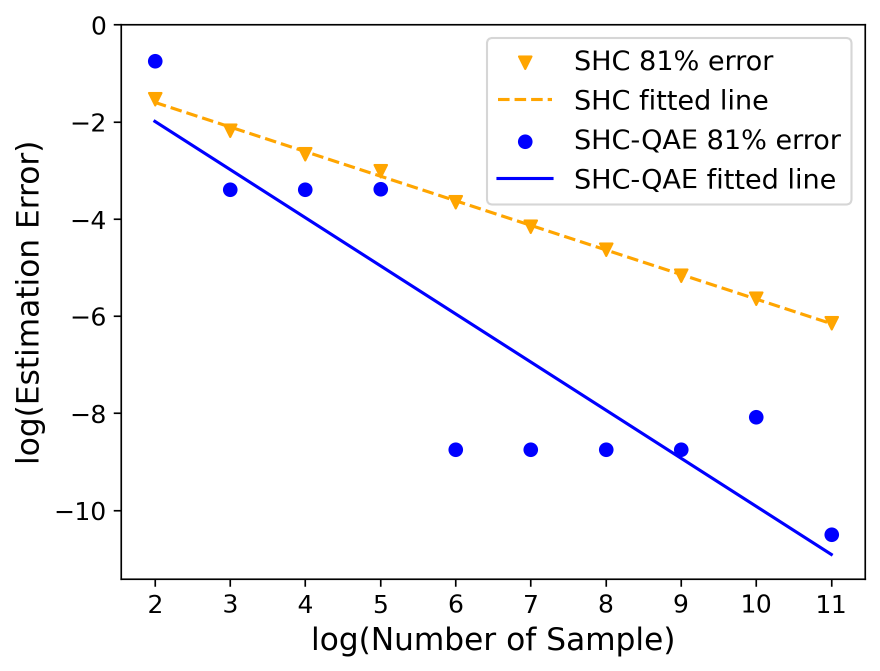}
 \caption{Linear fitting for SHC comparison}
\end{subfigure}
\begin{subfigure}{0.45\textwidth} 
 \centering
 \includegraphics[scale=0.4]{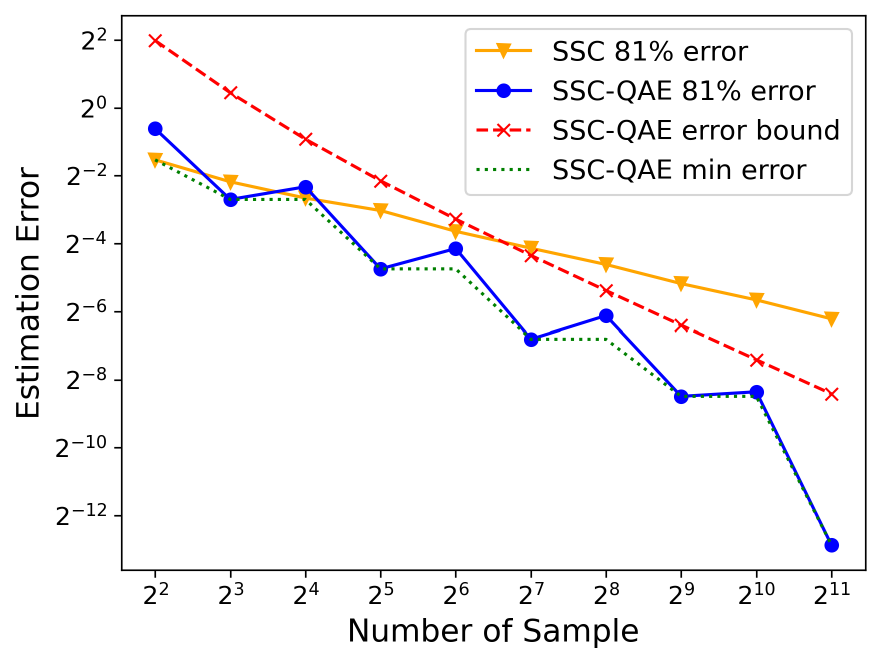}
 \caption{SSC comparison result}
\end{subfigure}
\begin{subfigure}{0.45\textwidth} 
 \centering
 \includegraphics[scale=0.4]{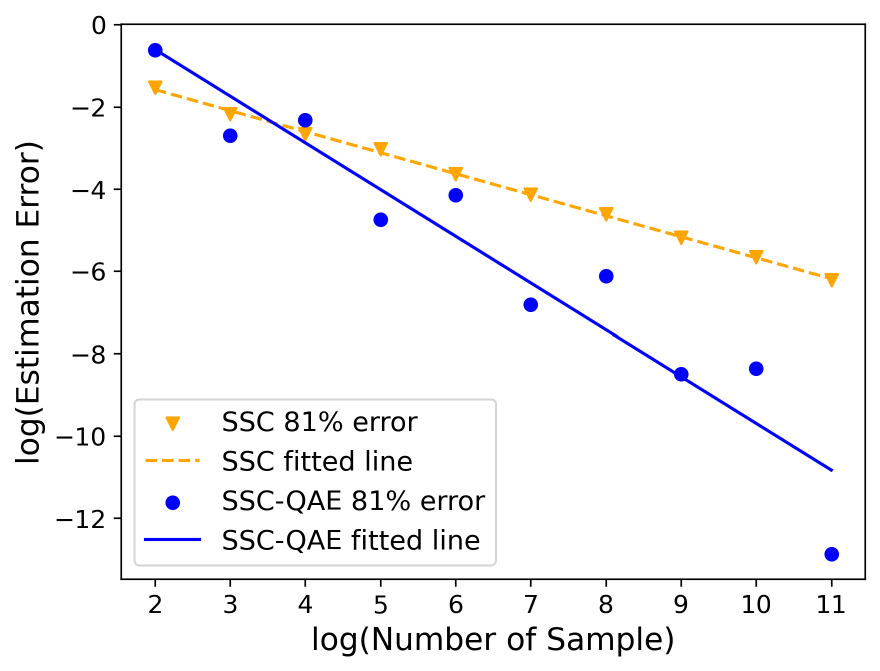}
 \caption{Linear fitting for SSC comparison}
\end{subfigure}
\caption{(a) The error scaling comparison result between SHC with QAE (SHC-QAE) and SHC with respect to the number of samples and (c) the comparison between SSC with QAE (SSC-QAE) and SSC. The answer ($a$) that each algorithm wants to estimate is $a\approx0.5952$ for SHC (also, for SHC-QAE) and $a\approx0.6541$ for SSC (also, for SSC-QAE). (b) and (d) is the linear fitting result of (a) and (c), respectively, using the $\log_2$ function. The slope ratio, the extent of speed-up, is calculated at approximately 1.957 for SHC and 2.221 for SSC.}
 \label{fig:SQKCs comparison}
\end{figure}

We conducted 2000 repetitions for both SQKCs-QAE and SQKCs circuits, each time using a specific number of samples. We sorted the errors in ascending order, and the 1621st value in the list corresponds to the $81\%$ error, which is plotted in blue curves with circles in Fig.~\ref{fig:SQKCs comparison}(a) and (c) for SHC and SSC, respectively. In these figures, the green dotted line correspond to the first element of the sorted list of SQKCs-QAE errors, indicating the minimum error. The dashed lines with $\times$ marker represent the upper bound of the estimation error in QAE, given by $2\sqrt{a(1-a)}\pi/N_q + \pi^2/N_q^2$, which QAE should satisfy with a probability of at least $81\%$ (see Eq.~(\ref{eq:qae error bound})).

Following numerical simulation, we fitted the estimation error results to straight lines in the log-log plot for comparing the sampling efficiency of SQKCs with and without QAE. The results of the linear fit are shown in Fig.~\ref{fig:SQKCs comparison}(b) and (d). The slope ratio for SHC and SSC are 1.957 and 2.221, respectively. Since the slope ratios of both SHC and SSC are approximately or exceeds 2, the linear fit results indicate that QAE indeed achieves quadratic speed-up in sampling, as expected.

Although the minimum and the 81$\%$ error for both SHC and SSC with QAE do not decrease monotonically with respect to the number of samples, they always remain below the theoretical upper bound of the estimation error in QAE. In essence, the deviation from the monotonic behavior (such as plateaus within certain ranges or the zig-zag pattern) can be attributed to the fact that we are estimating the continuous parameter $a$ via the discrete probability distribution. We elaborate further on this topic to justify such non-monotonic behavior in \ref{sec:AppendixB}.

\begin{figure}
 \centering
\begin{subfigure}{0.45\textwidth} 
 \centering
 \includegraphics[scale=0.4]{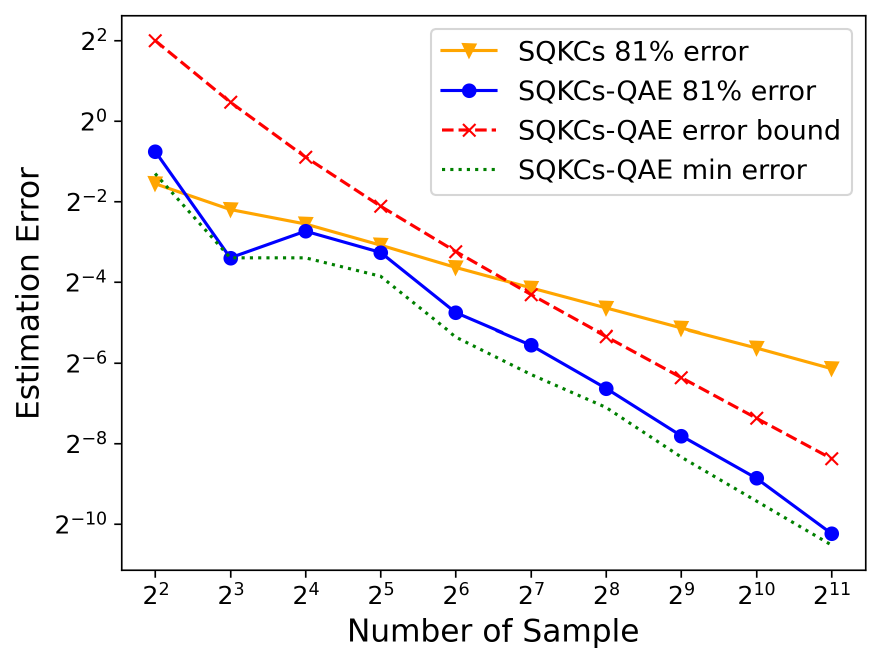}
 \caption{Average comparison result}
\end{subfigure}
\begin{subfigure}{0.45\textwidth} 
 \centering
 \includegraphics[scale=0.4]{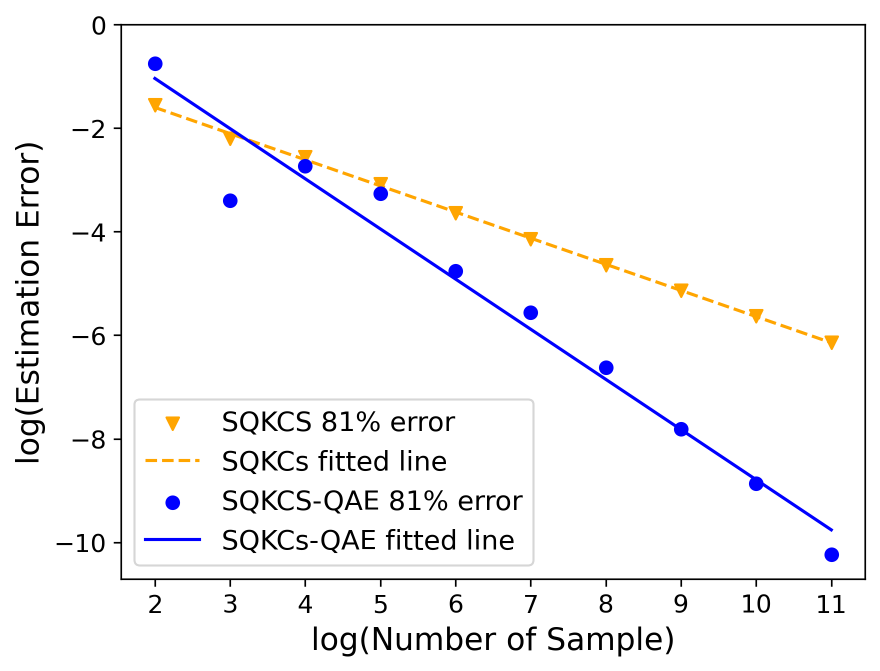}
 \caption{Linear fitting}
\end{subfigure}
 \caption{(a) Average error scaling comparison result between SQKCs-QAE and SQKCs. To obtain (a), we calculated an arithmetic mean for each error (also error bound) from a total of 12 comparison results, 6 for SHC and else for SSC. The data for each result include Eq. (\ref{eq:QAE data}) and 10 extra random data on the Iris dataset.
 (b) is the linear fitting result of (a), and the calculated slope ratio is approximately 1.9185.}
 \label{fig:SQKCs comparison average result}
\end{figure}

In addition, we calculate the mean estimation error, averaged across various values of $a$ determined by datasets different from the one specified in Eq.~(\ref{eq:QAE data}), as illustrated in Fig.~\ref{fig:SQKCs comparison average result}. 
This figure represents an average derived from 12 results, wherein 11 different datasets, including one same dataset Eq.~\ref{eq:QAE data} and 10 randomly selected training and test datasets, 5 each for both SHC and SSC, within the same features of the Iris dataset.
For detailed information on the computation of averages, refer to \ref{sec:data for average result}. The quadratic sampling advantage is also observed in this average result. Moreover, it presents a smoother error reduction curve with fewer plateaus or zig-zag patterns, supporting the idea that the non-monotonic behaviour depends on the specific value of the continuous parameter $a$. Furthermore, as this behaviour originates from estimating the continuous parameter through the discrete distribution, it can be mitigated by transforming the measurement results into a continuous distribution via techniques like maximum likelihood estimation (MLE). This approach will be presented in the subsequent section.

To conclude, based on the comparative simulation results between SQKCs-QAE and SQKCs in Fig.~\ref{fig:SQKCs comparison} and~\ref{fig:SQKCs comparison average result}, we validate that, for a given level of precision, the estimation speed of SQKCs can be quadratically enhanced by harnessing data superposition via QAE.

\subsubsection{Maximum likelihood QAE}
\label{sec:classifier with MLQAE}

Recent works have demonstrated alternative forms of QAE that do not rely on QPE and can significantly reduce the number of qubits and controlled gates needed compared to the traditional approach. Examples include Maximum Likelihood Quantum Amplitude Estimation (MLQAE)~\cite{suzuki2020amplitude} and other variants~\cite{grinko2021iterative, nakaji2020faster, manzano2023real}. Thus, by allocating saved resources from these variants to increase the number of training data, we can enhance the inherent performance of the classifiers while maintaining the same quadratic speed-up. More details on how the classification performance improves with the number of training data are provided in ~\ref{sec:multiple training data classifier}. Therefore, in this section, we apply MLQAE, which is the most suitable for our comparison method presented in Sec.~\ref{sec:method}, to SSC and verify it also can achieve the quadratic speed-up. 

\begin{figure}[t!]
\centering
\begin{minipage}[b]{.45\textwidth}\centering
\subfloat[SSC comparison result]{\includegraphics[scale=0.39]{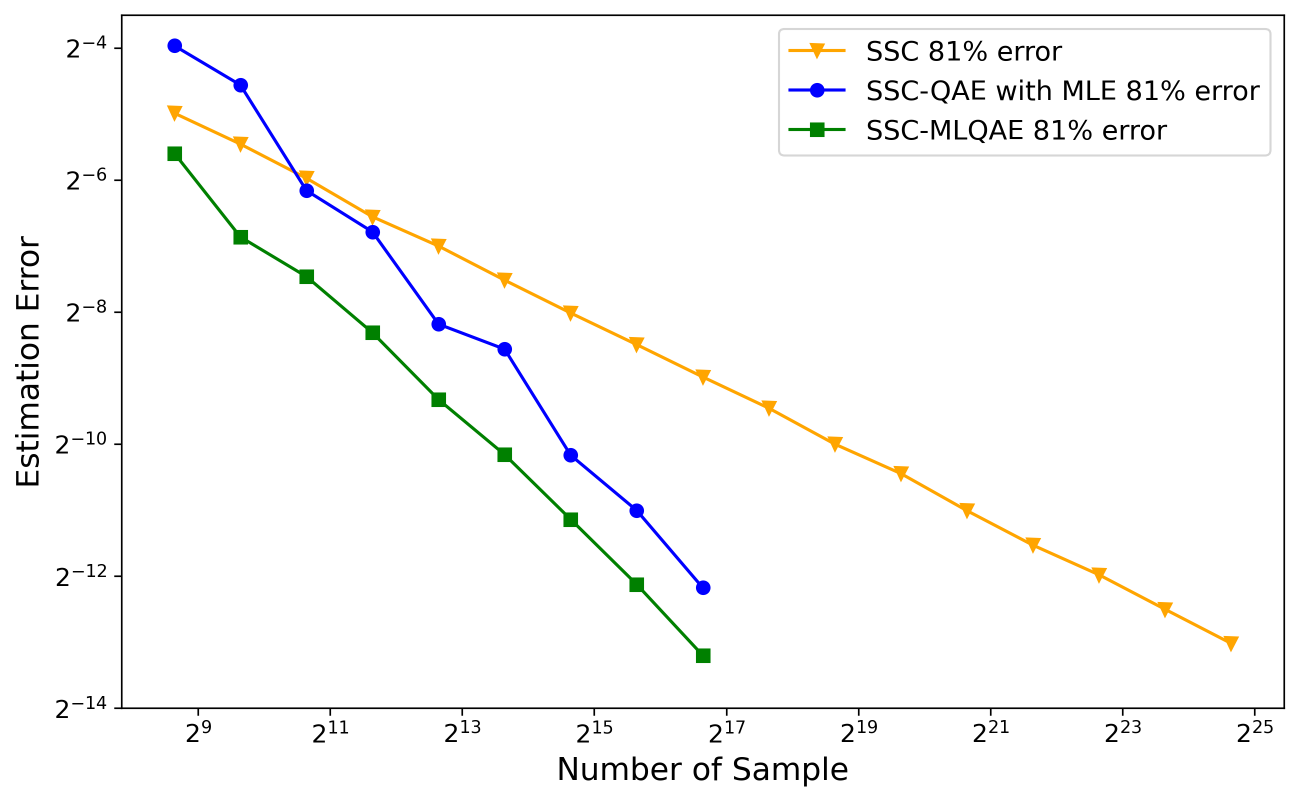}}
\end{minipage}\hfill
\begin{minipage}[b]{.45\textwidth}\centering
\includegraphics[scale=0.33]{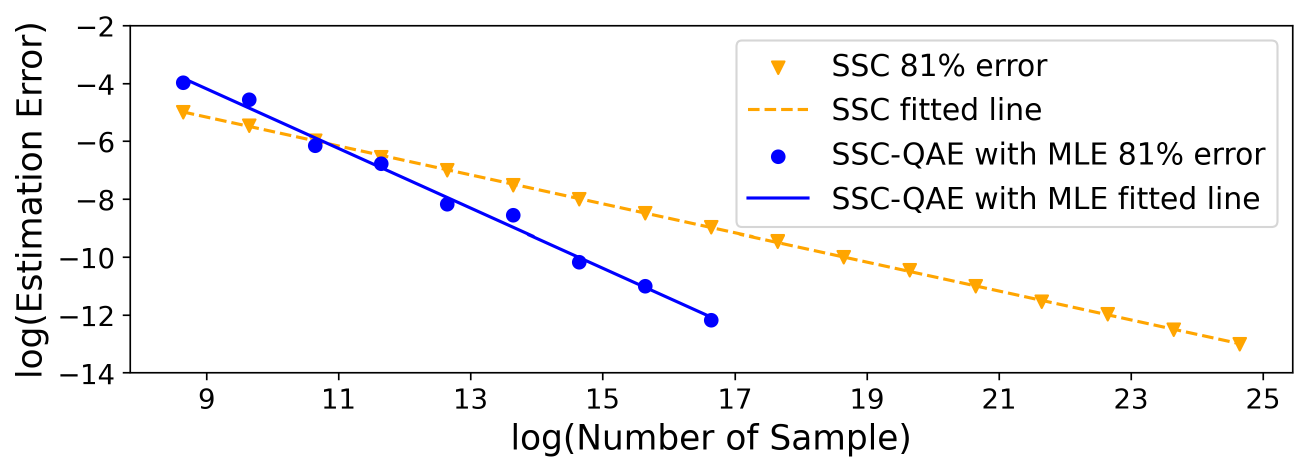}\\
\subfloat[Linear fitting]{\includegraphics[scale=0.33]{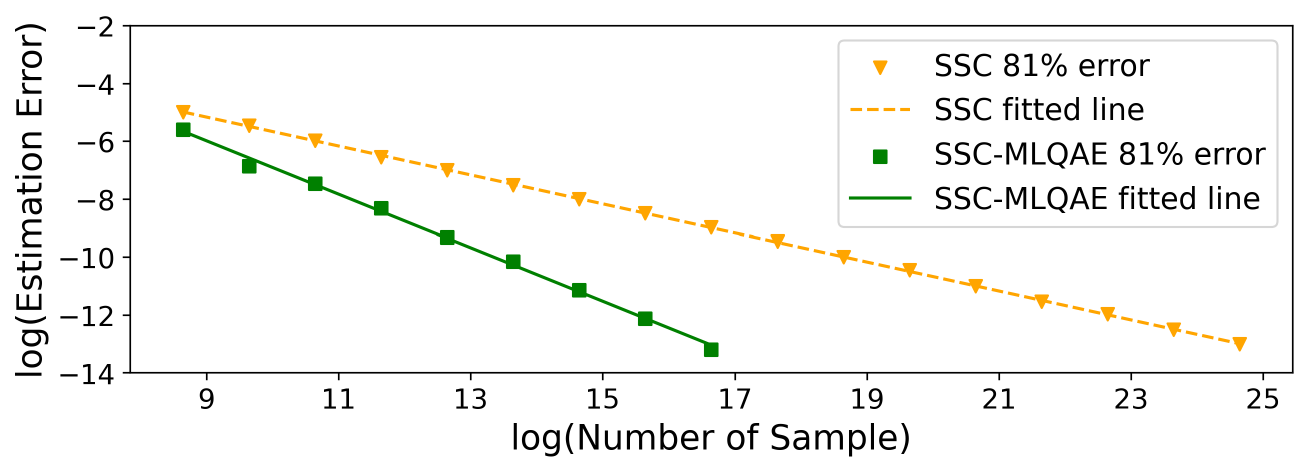}}
\end{minipage}
\caption{(a) The error scaling comparison result between SSC with MLQAE (SSC-MLQAE), SSC with QAE post-processed by MLE (SSC-QAE with MLE), and SSC. The answer ($a$) that each algorithm wants to estimate is $a\approx0.6541$. The linear fitting results of each case are shown in (b). The top figure of (b) represents the fitted line for SSC and  SSC-QAE with MLE, and the bottom is for SSC and QAE-MLQAE. The slope ratio, the extent of speed-up, is calculated at approximately 2.063 for SSC-QAE with MLE and 1.845 for SSC-MLQAE.}
\label{fig:SSC MLQAE result}
\end{figure}

The essential idea of MLQAE is to create a likelihood function from the measurements of several amplitude amplification processes instead of using QPE, requiring much fewer resources than the traditional QAE~\cite{suzuki2020amplitude}. Since MLQAE with an exponentially incremental sequence for the power of the Grover operator achieves an error of $\mathcal{O}(N^{-1}_q)$, we only consider this case of MLQAE. For a more explicit comparison, we also apply post-processing to the traditional QAE with MLE. QAE with MLE post-processing can be performed by applying MLE to the result distribution of the QAE. Applying MLE to QAE enables it to derive enhanced estimations and confidence intervals based on the likelihood ratio~\cite{grinko2021iterative}. This MLE post-processing converts the distribution of QAE into a continuous one, thereby improving the estimation and mitigating the non-monotonic behaviour observed in the previous section under traditional QAE, such as the zig-zag pattern and the long plateaus in the estimation error profile. 
The majority of comparison strategies are similar to Sec.~\ref{sec:method}. However, due to the nature of MLQAE and QAE with MLE requiring multiple shots to utilize MLE, we fixed the number of shots, $N^q_{shots}$, to 100. Thus, the number of samples becomes $2^{t+1}\times 100$, where $t$ is the number of ancilla qubits, and this becomes the number of shots for SSC, $N^c_{shot}$. We generated SSC with MLQAE (SSC-MLQAE), SSC with QAE post-processed by MLE (SSC-QAE with MLE), and SSC 1000 times for each sample size and displayed the 81st percentile value from multiple sorted errors for the result in Fig.~\ref{fig:SSC MLQAE result}. The set of data used in this simulation is identical to the one used in the simulations presented in Fig.~\ref{fig:SQKCs comparison}, which is the 23rd and 119th Iris samples for training data, and the 123rd Iris samples for test data. 

The slope ratio for SSC-QAE with MLE and SSC-MLQAE are approximately 2.063 and 1.845, respectively. Since the slope ratios of both approaches are close to 2, the linear fit results indicate these variants of QAE indeed achieve quadratic speed-up in sampling. It is important to note that SSC-MLQAE achieved the quadratic speed-up is achieved while using much smaller quantum circuits. As the size of the quantum circuit increases with the number of training data points in the superposition state, our result suggests that, for a fixed quantum circuit size, MLQAE has the potential to achieve higher classification accuracy by utilizing more training data points. 
Additionally, in the error scaling result of SSC-QAE with MLE, each point exhibits less deviation from the linearly fitted line. This observation further strengthens our argument that the plateaus and zig-zag patterns in Fig.~\ref{fig:SQKCs comparison} are attributed to the discrete nature of QAE.

\section{Conclusion}
\label{sec:conclusion}

Quantum computing opens exciting opportunities for the development of new machine learning approaches and methodologies. In connection with this, we outlined QKCs addressing the supervised binary classification problem. The potential advantage of QKCs stems from their ability to evaluate the classification score exponentially faster with respect to the number of features in the data, achieved by computing kernel using Hadamard or swap tests. However, as demonstrated in Sec.~\ref{sec:classification without superposition}, previous QKCs do not properly utilize the ability to superpose the entire dataset. Consequently, they are indistinguishable from evaluating the classification score by computing the kernel function independently for each training data point and aggregating them classically. The previous QKCs fail to attain any computational advantage from encoding the entire dataset as a quantum superposition state, despite the logarithmic increase in the number of qubits and the linear increase in circuit depth relative to the number of data samples. In response to this issue, we presented protocols that fully leverage the unique capability of quantum computers to superpose the entire training data, offering quadratic speed-up with respect to the number of training data via QAE. We also developed simplified versions of the QKCs (SQKCs), enabled by using the ordered encoding strategy. This results in a reduction of one qubit compared to previous QKCs and decreases the circuit depth linearly with respect to the number of training data. Additionally, we modified the measurement scheme to enable classification using a single-qubit projective measurement. This new measurement scheme is advantageous for practical integration of QAE into our SQKCs, as the target state of QAE can be marked with a single-qubit Pauli-$Z$ gate (i.e. the phase oracle). Finally, we demonstrated that the quantum speed-up persists in MLQAE~\cite{suzuki2020amplitude}, a variation of QAE that is more suitable for the near-term quantum devices.

The quantum classifiers discussed in this works are fully quantum in the sense that the classification score over the full dataset is computed entirely coherently on a quantum computer. On the other hand, QKM can solve classification problems in conjunction with classifier algorithms, such as SVM~\cite{havlivcek2019supervised,huang_power_2021,hur2023neural}. In such a hybrid scenario, a quantum computer is used only for constructing the kernel matrix, and the potential quantum advantage lies in the hardness of computing such kernel elements classically. Since the elements of the quantum kernel matrix are typically given by the fidelities between two quantum states that encodes two data points~\cite{havlivcek2019supervised}, the application of QAE can easily be extended to estimating them with quadratic speed-up.

While QKCs and SQKCs introduced in this paper focuses on binary problems, they can also address multi-class classification problems using heuristic strategies such as one-vs-rest or one-vs-one~\cite{GIUNTINI2023109956}. Furthermore, there is another interesting approach to making multi-class SSC inspired by its binary predecessor~\cite{Pillay2024}. Therefore, extending our method to multi-class QKCs could establish an interesting future work. Moreover, application of QAE in other approaches in QML, such as variational quantum algorithms, remains an interesting avenue for future research.

\section*{Acknowledgments}
This research was supported by Institute for Information \& communications Technology Promotion (IITP) grant funded by the Korea government (No. 2019-0-00003, Research and Development of Core technologies for Programming, Running, Implementing and Validating of Fault-Tolerant Quantum Computing System), by the National Research Foundation of Korea (Grant Numbers: 2022M3E4A1074591, 2023M3K5A1094805, 2023M3K5A1094813), by the KIST Institutional Program (2E32941-24-008), and by the Yonsei University Research Fund of 2023 (2023-22-0072).


\newpage
\appendix
\setcounter{section}{0}
\renewcommand{\thesection}{\text{Appendix }\Alph{section}} 
\renewcommand\thefigure{\Alph{section}.\arabic{figure}} 
\renewcommand\theequation{\Alph{section}.\arabic{equation}} 
\renewcommand\thetable{\Alph{section}.\arabic{table}} 

\section{Calculating classification score without data superposition}
\label{sec:weighted Hadamard and swap test}
\setcounter{figure}{0}
\setcounter{equation}{0}

Recall that the mixed state obtained from the partial trace of the index register in QKCs is $\rho(x_m,\tilde{x},y_m)=\sum_{m=0}^{M-1}w_m |\psi^l(m)\rangle\langle\psi^l(m)|\otimes |y_m\rangle\langle y_m|$.  
We mentioned in the main manuscript that performing weighted Hadamard or swap test on the state $|\psi^l(m)\rangle$ individually and aggregating the total of $M$ outcomes classically can yield the same classification score without data superposition. In this section, we explicitly demonstrate the unnecessity of superposing the training data points in the previous HC and SC by introducing the weighted Hadamard test (WHT) and weighted swap test (WST) and constructing classifiers with them. 

\begin{figure}[htb!]
  \centering
\begin{subfigure}{0.45\textwidth} 
  \centering
  \includegraphics[scale=0.4]{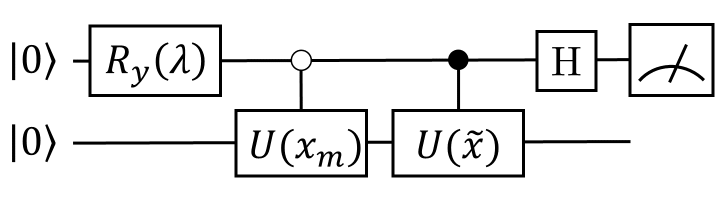}
  \caption{Weighted Hadamard test}
  \label{fig:Weighted H test}
\end{subfigure}
\begin{subfigure}{0.45\textwidth} 
  \centering
  \includegraphics[scale=0.4]{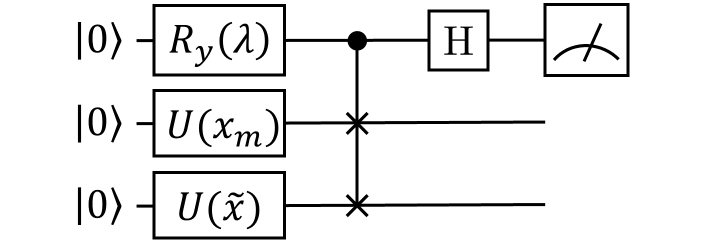}
  \caption{Weighted swap test}
  \label{fig:Weighted S test}
\end{subfigure}
  \caption{Quantum circuit diagrams for (a) weighted Hadamard test (WHT) and (b) weighted swap test (WST), used to implement the QKCs without data superposition. By aggregating the expectation values of the circuit for each training data point, the QKCs can be constructed without data superposition.}
  \label{fig:H and S test}
\end{figure}

The quantum circuit diagrams for WHT and WST are depicted in Fig.~\ref{fig:H and S test}, where $U(x_m)\text{ and }U(\tilde{x})$ indicate the data encoding (e.g. amplitude encoding) operator for $m$th training data and test data to initialize them as $|x_m\rangle$ and $|\tilde{x}\rangle$, respectively. The output of both circuits can be described by the expectation value of the one-qubit observable as, $\langle \sigma_z^l\rangle_m = \sin(\lambda) k(x_m,\tilde{x})$, where $l\in\lbrace h,s\rbrace$ is the label for indicating whether the state is for WHT or WST and $k(x_m, \tilde{x})$ is the kernel computed as $\Re\left( \langle x_m|\tilde{x}\rangle\right)$ for WHT and  $\big|\langle x_m|\tilde{x}\rangle\big|^2$ for WST. For supervised classification, one can choose $\lambda$ such that $\sin(\lambda) =(-1)^{y_m}w_m$. 
Therefore, each circuit can obtain the weighted kernel of one training data, $\langle \sigma_z^l\rangle_m=(-1)^{y_m}w_m k(x_m,\tilde{x})$, so that if we aggregate every expectation value for $m$th circuit, $\langle \sigma_z^l\rangle_m$, we can calculate the classification score for $M$ number of training data as 
\begin{equation}
  \langle\sigma_z^l\rangle = \sum^{M-1}_{m=0} \langle \sigma_z^l\rangle_m = \sum^{M-1}_{m=0}(-1)^{y_m}w_m k(x_m,\tilde{x}).
\end{equation}
It is the same measure as Eqs.~(\ref{eq:HC two measurement observable}) and (\ref{eq:SC two measurement observable}), which can be gained from HC and SC, respectively. Consequently, the classification score of the QKCs can also be evaluated by computing the kernel function for the test data and each training data independently via WHT or WST and aggregating every expectation value classically.

\section{Interpretation of the QAE results}
\label{sec:AppendixB}
\setcounter{figure}{0}

In this section, we elaborate on the theoretically expected nature of the two non-monotonic behaviors observed in the minimum and the $81\%$ error of QAE with respect to the sample size, as shown in Fig.~\ref{fig:SQKCs comparison}.

The first behavior, referred to as plateaus in the text, describes instances where the estimation error of QAE remains constant within a specific range, instead of monotonically decreasing with the sample size. This phenomenon is primarily observed in Fig.~\ref{fig:SQKCs comparison}(a). For example, solely considering the results after $2^6$ samples in the figure might be misleading, because comparing the slope of the linear fit using only those points in that flat region could suggest that the performance of SHC-QAE is worse than that of SHC without QAE. Such flat regions can emerge because QAE estimates the continuous value $a$ using a discrete number of bits, and in some cases, the addition of a few more bits may not necessarily enhance the approximation. For example, when aiming to estimate $a = 0.01$ using QAE, the estimation error remains constant when using between one and four ancilla qubits. This is because, among the values that QAE with four ancilla qubits can estimate, the smallest non-zero value is 0.0381, calculated as $\sin(\pi/2^4)^2$. Moreover, with fewer ancilla qubits, this value would be even larger. Therefore, the QAE outputs an estimate $\tilde{a}=0$, since this is a better estimate of $a$ than $0.0381$, until using more than four ancilla qubits. With five ancilla qubits, the smallest non-zero value that QAE can output is 0.0096, which is a better estimate than 0. In other words, the minimum error of QAE with up to four ancilla qubits is 0.01, resulting in the plateau. However, this minimum error reduces to 0.0004 when the number of ancilla is five. It is important to note that even though the flat region exists, all estimation errors are below the QAE error bound and agree with the theory.
\begin{figure}[h!]
\centering
 \begin{subfigure}[b]{0.45\textwidth}
  \includegraphics[scale=0.45]{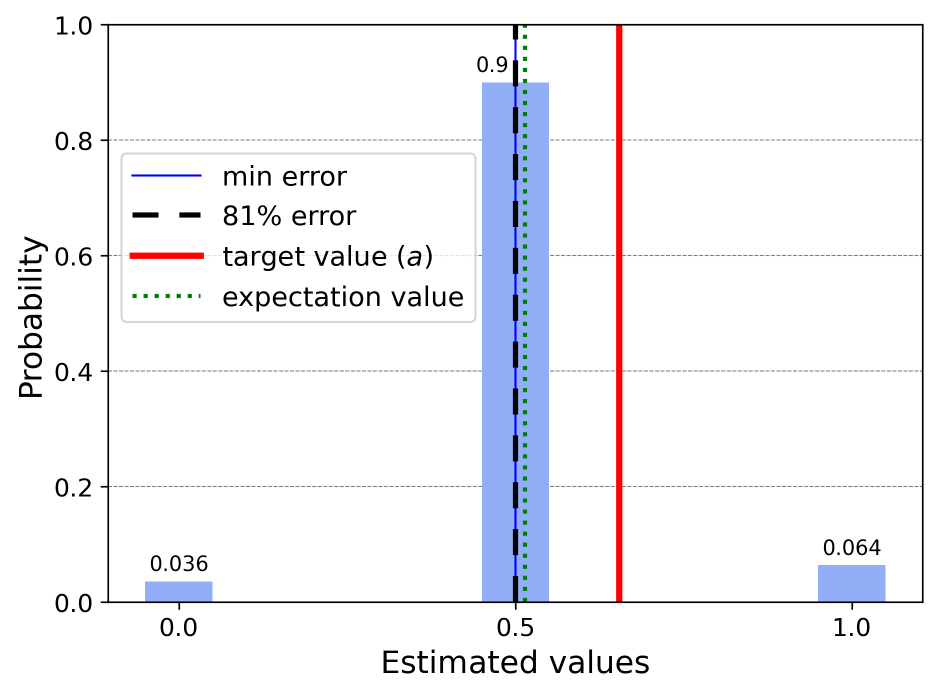}
  \caption{SSC-QAE result distribution for two ancilla}
 \end{subfigure}
 \begin{subfigure}[b]{0.45\textwidth}
  \includegraphics[scale=0.45]{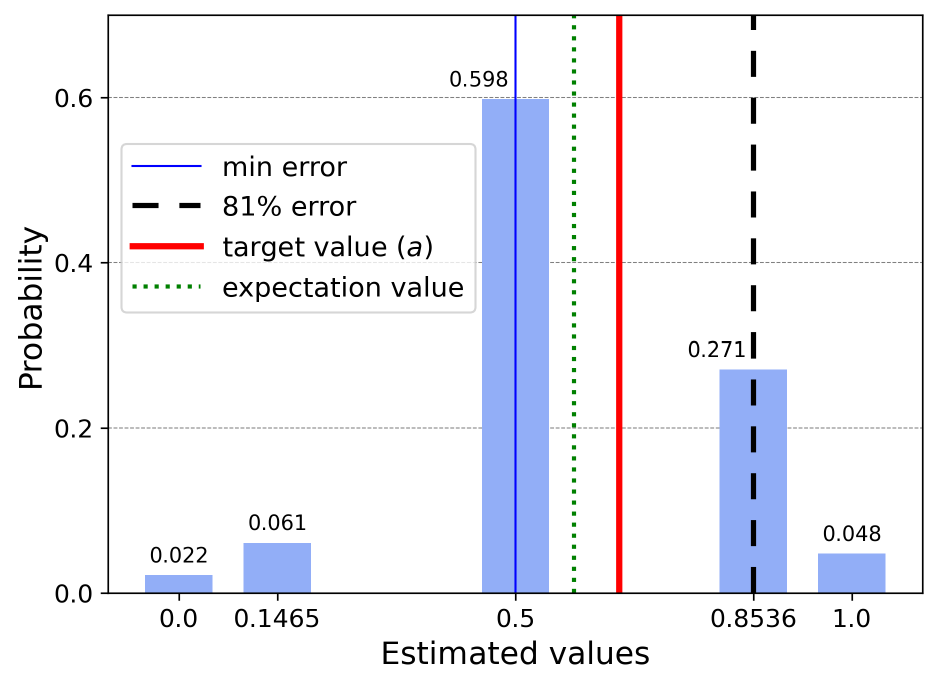}
  \caption{SSC-QAE result distribution for three ancilla}
 \end{subfigure}
\caption{Probability distribution of (a) two-ancilla SSC-QAE result and (b) three-ancilla SSC-QAE result to estimate the value $a \approx 0.6541$, which is represented in red ticker solid vertical line. In both cases, the QAE result indicating the minimum error is 0.5 shown in blue thin solid line and its derivation probability is 0.9 and 0.598 in (a) and (b), respectively. Therefore, in two-ancilla QAE, the results associated with an 81\% error, black dashed line, is 0.5. In contrast, it is 0.8536 in three-ancilla QAE. The green dotted vertical line represents the expectation value of the probability distribution. It is closer to $a$ in three-ancilla QAE than in two-ancilla QAE, indicating that three-ancilla QAE performs better in predicting $a$ than two-ancilla QAE.}
\label{fig:SSC-QAE result distribution}
\end{figure}

The second issue, primarily observed in Fig.~\ref{fig:SQKCs comparison}(c), is the zig-zag shape of the SSC-QAE $81\%$ error, unlike the monotonic decrease of the SSC $81\%$ error with the sample size. While the minimum SSC-QAE error remains constant, several jumps in the estimation error are observed in the QAE 81 percentile error. This zig-zag pattern arises due to the fact that the resolution of the discrete probability distribution corresponding to the QAE result improves with the number of samples, and that the 81 percentile largest error is selected from such a discrete distribution.

For example, Fig~\ref{fig:SSC-QAE result distribution} shows the probability distribution of SSC-QAE results depicted in Fig.~\ref{fig:SQKCs comparison}(c), especially for cases using two and three ancilla qubits, respectively. In this example, the true answer is $a=0.6541$. In the two-ancilla QAE, the resolution of the discrete probability distribution is low, such that it can only output 0, 0.5, or 1. Although the true value lies between 0.5 and 1, the probability to obtain 1 is very small, whereas the probability to output 0.5 is approximately 0.9. Therefore, the $81$st largest value among all errors would likely be the difference between $\vert 0.5-a\vert = 0.1541$. On the other hand, the resolution improves in the three-ancilla QAE, enabling it to output 0, 0.1465, 0.5, 0.8536, or 1. The true value lies between 0.5 and 0.8536. There is now approximately a 0.27 probability of obtaining 0.8536. Thus, the $81$st largest value among all errors would likely be $\vert 0.8536-a\vert = 0.1995$; the 81 percentile error has increased as $t$ increases from 2 to 3. However, the most frequent outcome in both two- and three-ancilla QAE are 0.5, which yields the minimum error in both cases. Again, it is important to note that even though the zig-zag pattern is observed, all estimation errors are below the QAE error bound and agree with the theory. As a side note, the expected outputs of two-ancilla QAE and three-qubit QAE are (i.e. expectation values of the probability distributions in Fig~\ref{fig:SSC-QAE result distribution}) 0.5140 and 0.5873, respectively. Thus, the expected estimation improves as $t$ increases from 2 to 3.

\section{Data for average results}
\setcounter{figure}{0}
\setcounter{table}{0}
\label{sec:data for average result}

\begin{table}[htb!]
\setlength{\tabcolsep}{7pt}
\renewcommand{\arraystretch}{1.5}
\centering
\begin{subtable}{\textwidth}\centering
\vspace{-0.3cm}
\begin{tabular}{|c|c|c|c|}
\hline
$|\tilde{x}\rangle$ & $0.3856|0\rangle + 0.9227|1\rangle$ (123) & $0.3436|0\rangle + 0.9391|1\rangle$ (33)  & $0.8882|0\rangle + 0.4594|1\rangle$ (130)  \\
$|x_0\rangle$       & $0.9635|0\rangle + 0.2676|1\rangle$ (23)  & $0.3162|0\rangle + 0.9487|1\rangle$ (17)  & $0.3714|0\rangle + 0.9285|1\rangle$ (1)  \\
$|x_1\rangle$       & $0.3526|0\rangle + 0.9358|1\rangle$ (119) & $0.8882|0\rangle + 0.4594|1\rangle$ (105) & $0.8914|0\rangle + 0.4532|1\rangle$ (103) \\
 $a$ & 0.5952&  0.4343&  0.4391\\
\hline
$|\tilde{x}\rangle$ & $0.4158|0\rangle + 0.9095|1\rangle$ (44)  & $0.3757|0\rangle + 0.9267|1\rangle$ (22)  & $0.3443|0\rangle + 0.9389|1\rangle$ (14) \\
$|x_0\rangle$       & $0.4229|0\rangle + 0.9062|1\rangle$ (13)  & $0.4356|0\rangle + 0.9002|1\rangle$ (10)  & $0.4472|0\rangle + 0.8944|1\rangle$ (21)  \\
$|x_1\rangle$       & $0.8638|0\rangle + 0.5039|1\rangle$ (127) & $0.9358|0\rangle + 0.3526|1\rangle$ (119) & $0.8838|0\rangle + 0.4679|1\rangle$ (102) \\
 $a$ &  0.5456&  0.5799&  0.4375\\
\hline
\end{tabular}
\caption{Data used in SHC-QAE vs SHC experiments}
\vspace{0.3cm}
\end{subtable}
\begin{subtable}{\textwidth}\centering
\begin{tabular}{|c|c|c|c|}
\hline
$|\tilde{x}\rangle$ &$0.3856|0\rangle + 0.9227|1\rangle$ (123)& $0.8944|0\rangle + 0.4472|1\rangle$ (129) & $0.3482|0\rangle + 0.9374|1\rangle$ (37)  \\
$|x_0\rangle$       &$0.9635|0\rangle + 0.2676|1\rangle$ (23)& $0.3162|0\rangle + 0.9487|1\rangle$ (34)  & $0.4258|0\rangle + 0.9048|1\rangle$ (27)  \\
$|x_1\rangle$       &$0.3526|0\rangle + 0.9358|1\rangle$ (119)& $0.8882|0\rangle + 0.4594|1\rangle$ (105) & $0.8973|0\rangle + 0.4413|1\rangle$ (136) \\
$a$ &  0.6541&  0.6250&0.6164\\
\hline
$|\tilde{x}\rangle$ & $0.8779|0\rangle + 0.4789|1\rangle$ (117) & $0.8662|0\rangle + 0.4997|1\rangle$ (148) & $0.8944|0\rangle + 0.4472|1\rangle$ (133) \\
$|x_0\rangle$       & $0.3482|0\rangle + 0.9374|1\rangle$ (41)  & $0.3162|0\rangle + 0.9487|1\rangle$ (17)  & $0.3511|0\rangle + 0.9363|1\rangle$ (36)  \\
$|x_1\rangle$       & $0.8720|0\rangle + 0.4895|1\rangle$ (121) & $0.8720|0\rangle + 0.4895|1\rangle$ (121) & $0.8838|0\rangle + 0.4679|1\rangle$ (143) \\
$a$ &  0.3924&  0.6101&0.3844\\
\hline
\end{tabular}
\caption{Data used in SSC-QAE vs SSC experiments}
\end{subtable}

\caption{Data of (a) SHC-QAE vs SHC experiments and (b) SSC-QAE vs SSC experiments each depicted in Figs.~\ref{fig:H average data} and \ref{fig:S average data}. $|\tilde{x}\rangle \text{ and } |x_m\rangle$ is test data and training data, respectively, where subscript $m$ indicates the class of the training data and the number within parentheses represents the index of each data in the Iris sample.}
\label{table:data for avg}
\end{table}

\begin{figure}[hbt!]
\hspace{-0.46cm}
\centering
\vspace{-0.15cm}
  \begin{tabular}{cc}
    \includegraphics[width=0.3\textwidth]{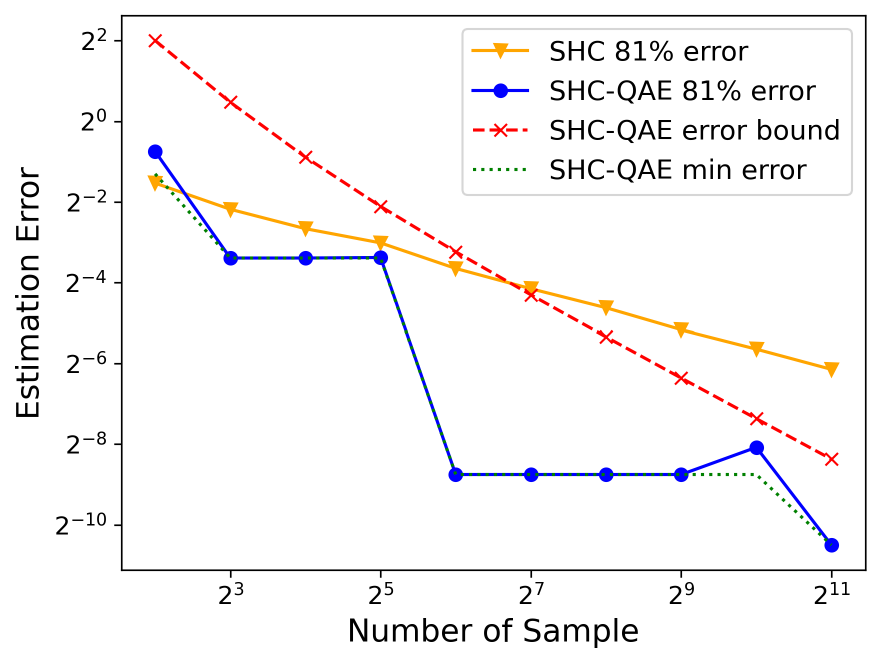}
    \includegraphics[width=0.3\textwidth]{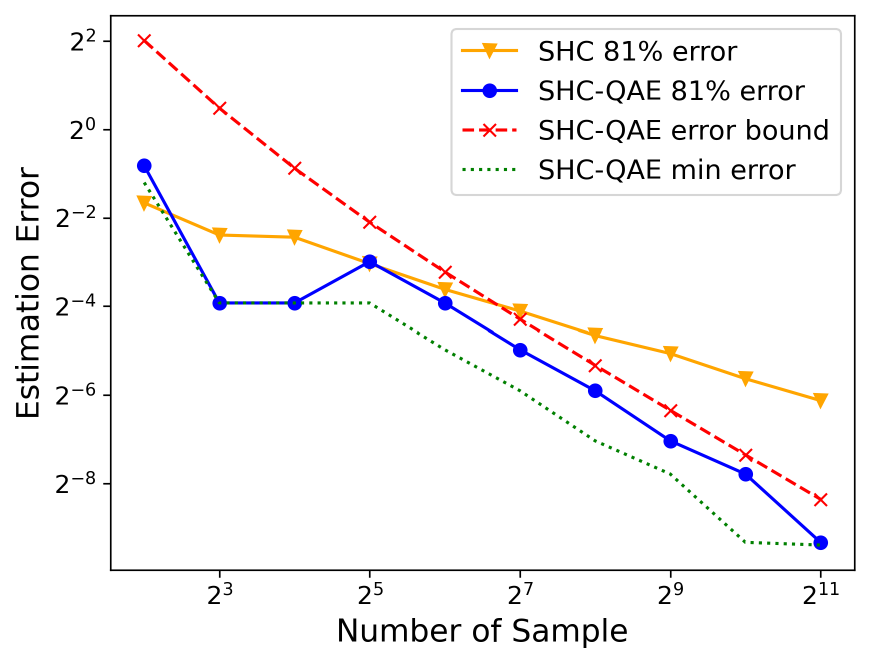}
    \includegraphics[width=0.3\textwidth]{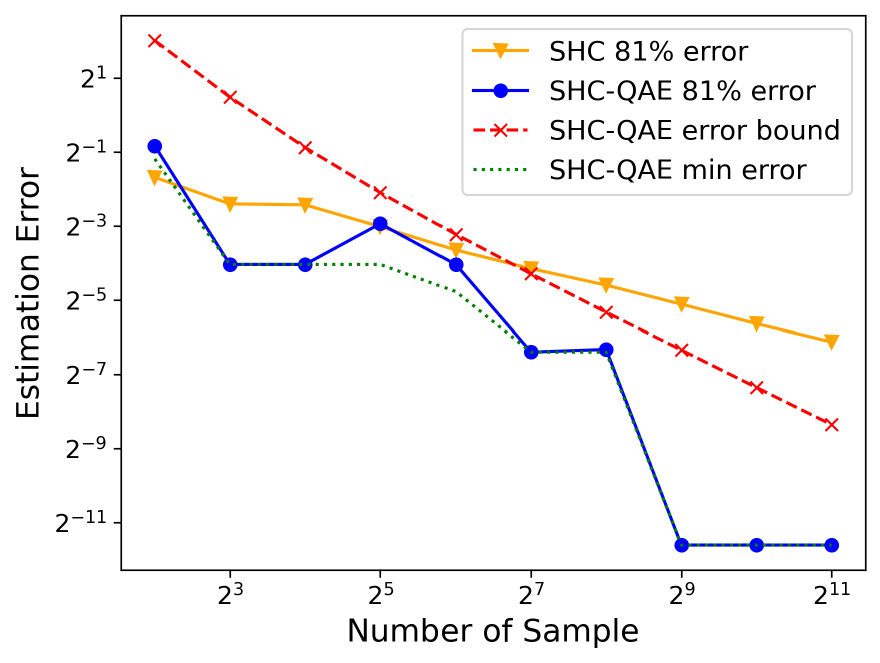}\\
    \includegraphics[width=0.3\textwidth]{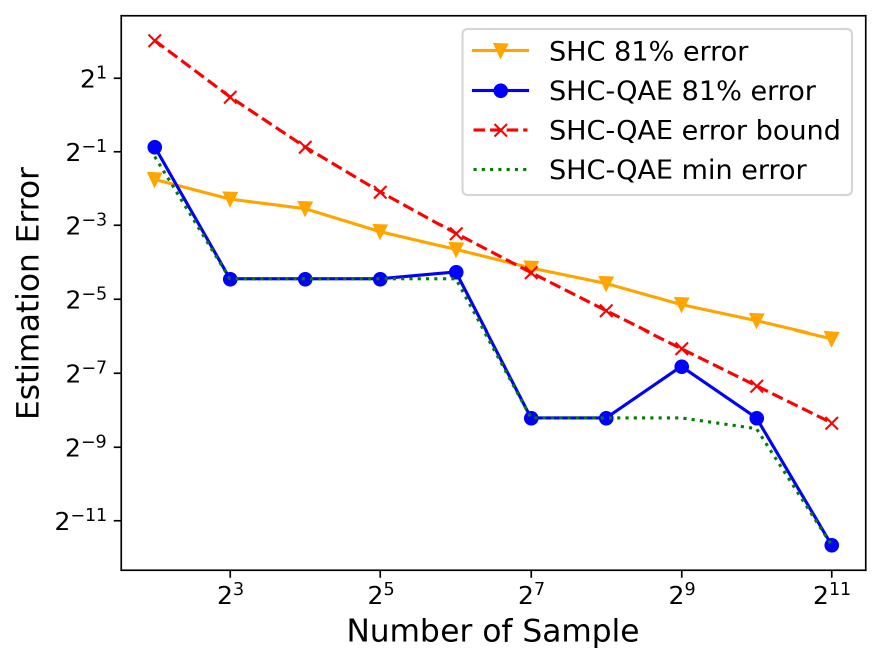}
    \includegraphics[width=0.3\textwidth]{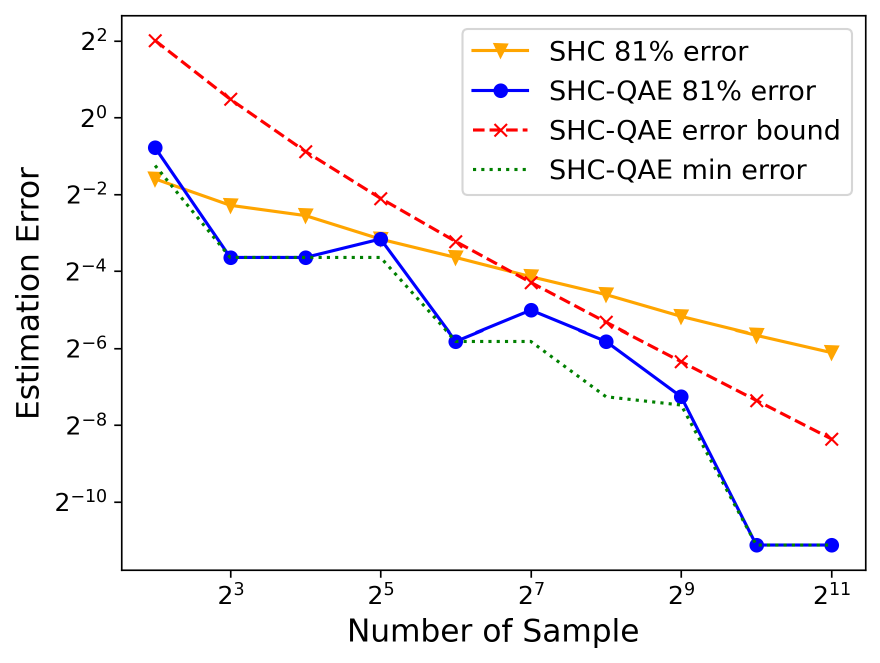}
    \includegraphics[width=0.3\textwidth]{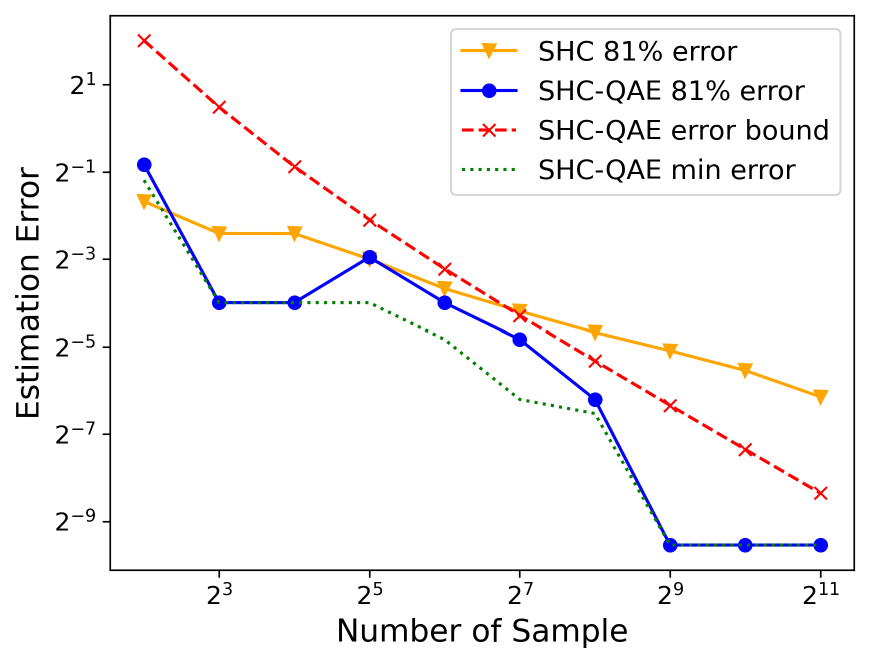}\\    
  \end{tabular}
  \caption{The error scaling comparison result between SHC-QAE and SHC on 6 different data (the left top one is for the dataset shown in Eq.~(\ref{eq:QAE data}) and the rest are random data which are organized in Table~\ref{table:data for avg}(a)) for calculating the average result in the figure~\ref{fig:SQKCs comparison average result}.}
  \label{fig:H average data}
\end{figure}

\begin{figure}[hbt!]
\hspace{-0.46cm}
\centering
\vspace{-0.15cm}
  \begin{tabular}{cc}
    \includegraphics[width=0.3\textwidth]{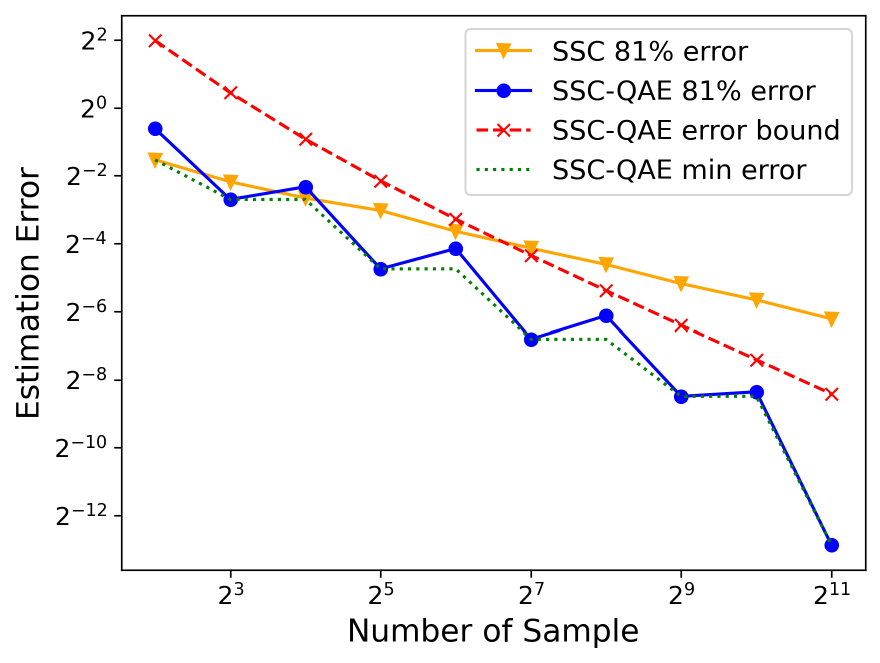} 
    \includegraphics[width=0.3\textwidth]{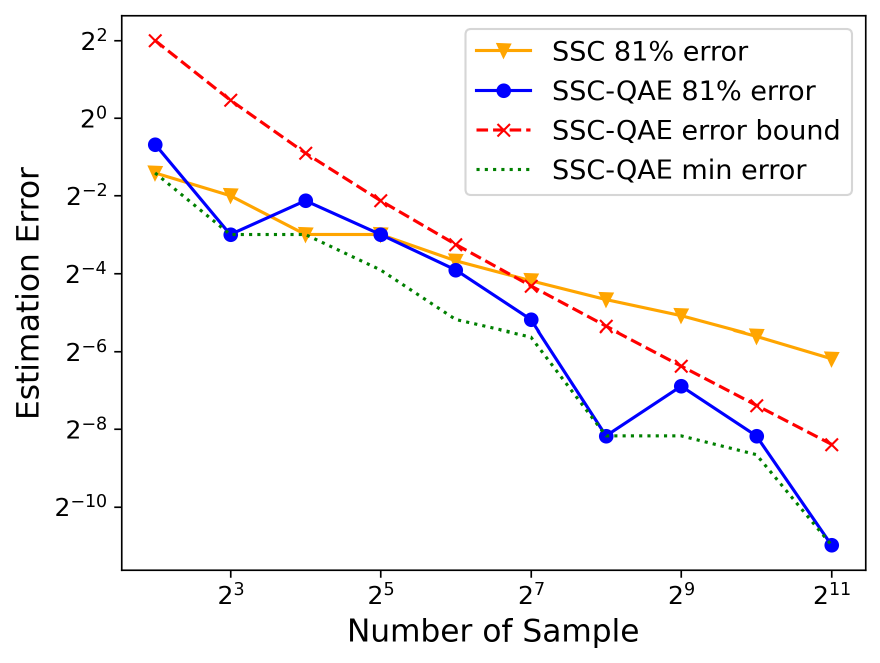}
    \includegraphics[width=0.3\textwidth]{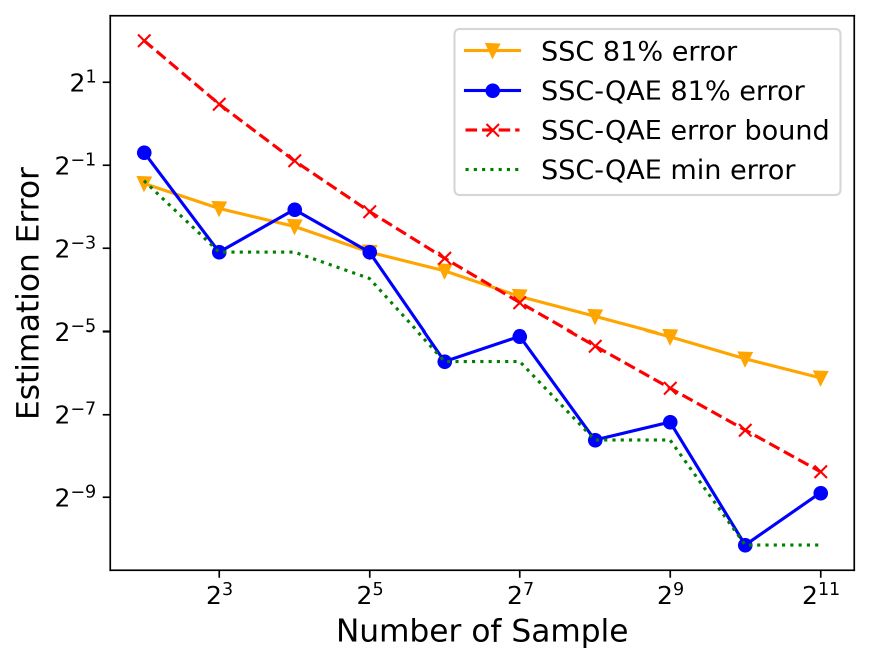}\\
    \includegraphics[width=0.3\textwidth]{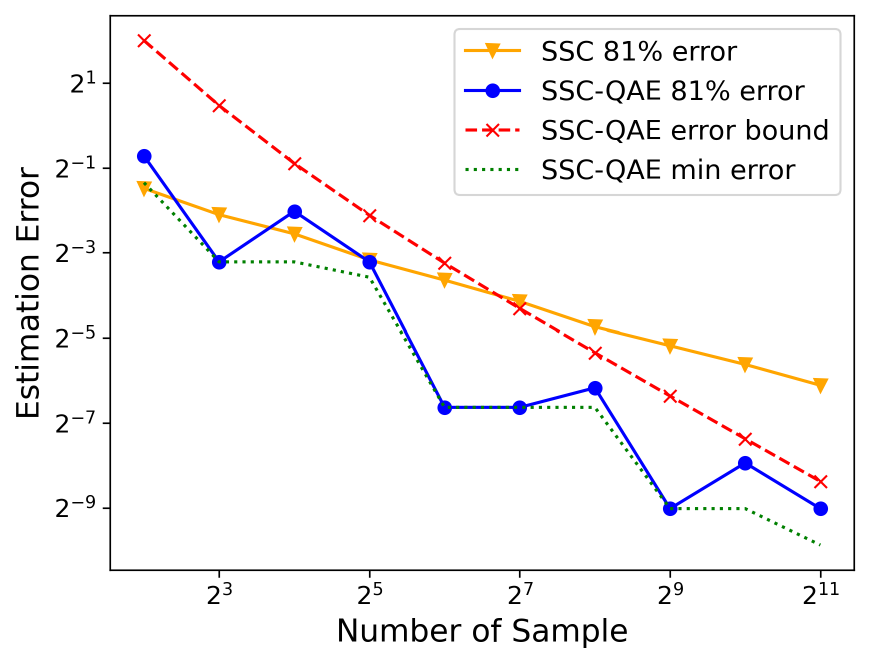}
    \includegraphics[width=0.3\textwidth]{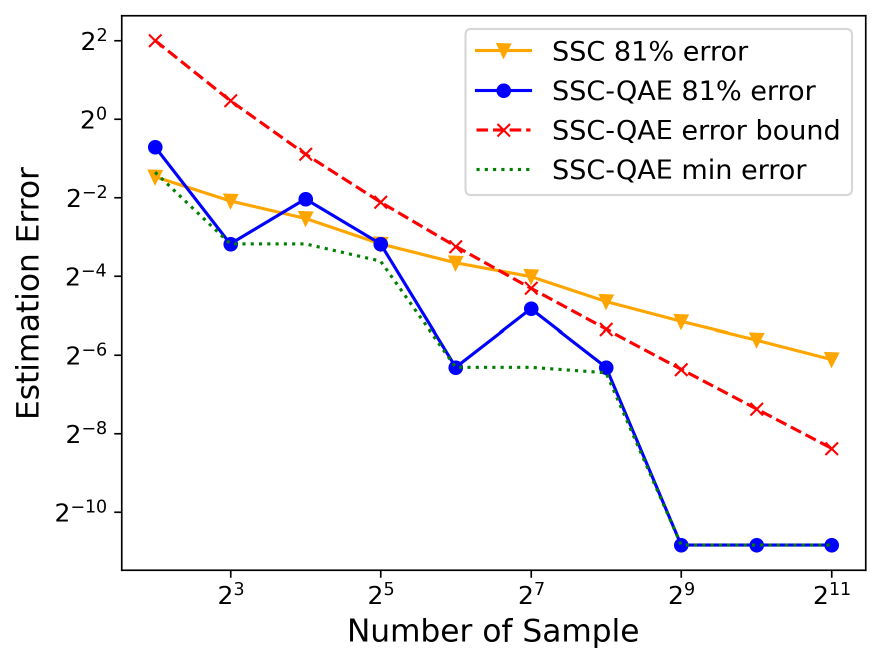}
    \includegraphics[width=0.3\textwidth]{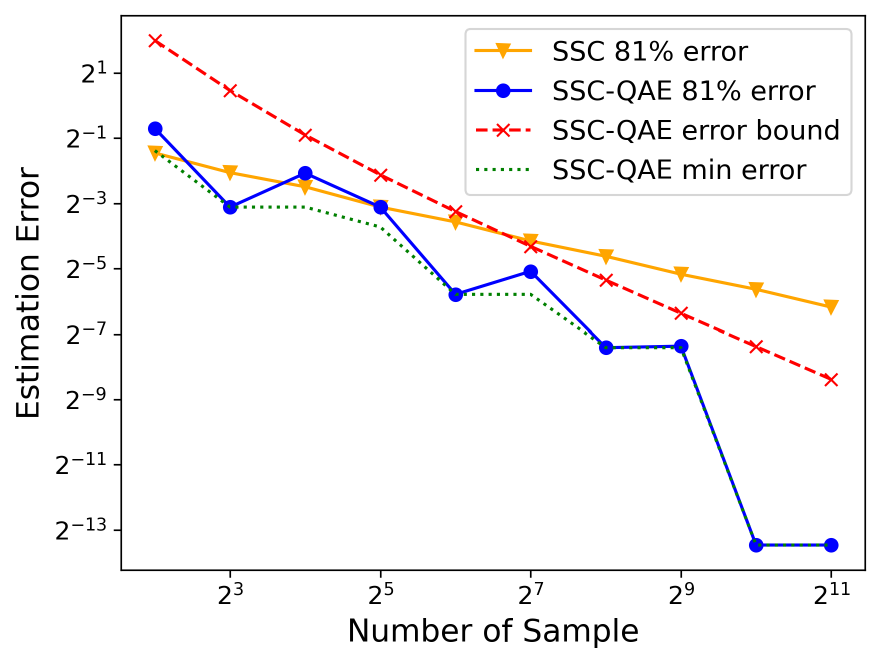}\\
  \end{tabular}
  \caption{The error scaling comparison result between SSC-QAE and SSC on 6 different data (the left top one is for  the dataset shown in Eq.~(\ref{eq:QAE data}) and else are random data which are organized in Table~\ref{table:data for avg}(b)) for calculating the average result in the figure~\ref{fig:SQKCs comparison average result}.}
  \label{fig:S average data}
\end{figure}

We calculate the arithmetic mean of 12 comparison results, 6 for the SHC-QAE vc SHC and else for the SSC-QAE vs SSC, in the manuscript to explore the average results. The Figs.~\ref{fig:H average data} and~\ref{fig:S average data} are specific comparison results used in Fig.~\ref{fig:SQKCs comparison average result}, and among them, the left top data is the dataset that we already treated in Sec.~\ref{sec:result}, and others are the results of random test and training dataset in the Iris data set, where classes : \textit{setosa} and \textit{virginica},  features : \textit{sepal width} and \textit{petal length}. The random dataset used in Figs.~\ref{fig:H average data} and \ref{fig:S average data} is organized in Table~ \ref{table:data for avg}.

\section{Classifier inherent error}
\setcounter{figure}{0}
\label{sec:multiple training data classifier}

 \begin{figure}[htb!]
\centering
 \begin{subfigure}[b]{0.45\textwidth}
  \includegraphics[scale=0.45]{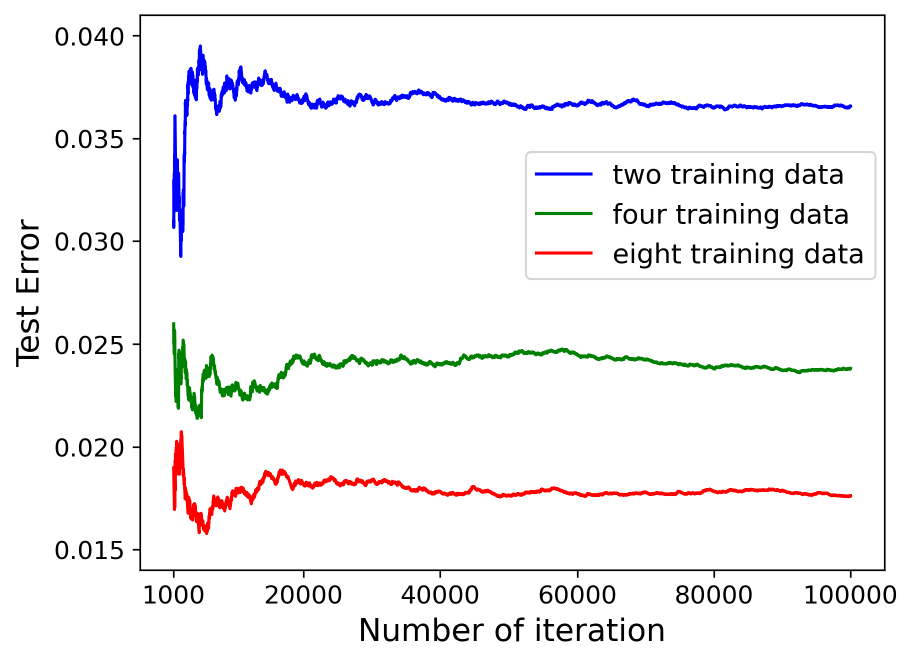}
  \caption{SHC error}
  \label{fig:SHC error}
 \end{subfigure}
 \begin{subfigure}[b]{0.45\textwidth}
  \includegraphics[scale=0.45]{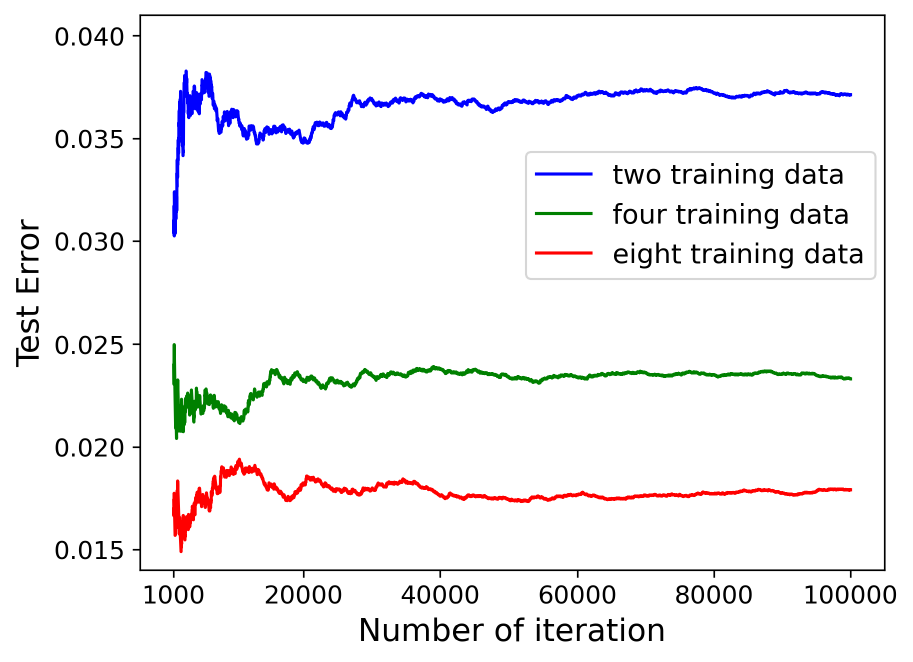}
  \caption{SSC error}
  \label{fig:SSC error}
 \end{subfigure}
\caption{Classifier inherent error for two (one pair), four (two pairs), and eight (four pairs) training data. In both SHC and SSC, the test error converges after multiple iterations to approximately 0.037, 0.024, and 0.018, respectively, when there are 2, 4, and 8 training data.}
\label{fig:classifier error}
\end{figure}
Figure~\ref{fig:classifier error} shows error convergences between three different numbers of data in both classifiers, respectively. We iterated the classifiers 100000 times for random data and accumulated the number of wrong classifications plotting the error rate each 10 iterations after the first 1000 iterations. 
For the random data, we used the first two features and classes in the IRIS dataset, classes : \textit{setosa} and \textit{versicolor}, features : \textit{sepal length} and \textit{sepal width}. 
The reason why we used different features of data from the main text is that the two features used in Sec.~\ref{sec:result} had a 0.00 classification error for the two training data in both classifiers since the dataset is perfectly separable. However, note that it does not imply that our algorithms did not require data pre-processing on account of this fact. The original HC used the post-selection scheme, and standardizing data to a zero mean and unit standard deviation was necessary to elevate the probability of selecting the appropriate branch to around 1/2. 
The effectiveness of SQKCs for the dataset used here can be significantly enhanced by implementing feature maps, such as polynomial feature maps~\cite{suthaharan2016support}, to the data. Additionally, it can also be improved by increasing the number of training data while maintaining the full quantumness of the classification. According to the results, the convergence error is approximately 1.3 to 1.5 times smaller when the amount of training data is doubled for both classifiers. In other words, if the number of data points increases by a factor of four, the errors decrease by a factor of two. However, since the rate at which errors are reduced appears to be gradually decreasing, it would be inappropriate to claim that increasing the number of training data is unconditionally efficacious. Nevertheless, it is evident that using more training data can effectively reduce error rates. Therefore, adapting variants of QAE, such as MLQAE, which allow for the utilization of more data within the constraints of a given circuit size, can be beneficial not only for executing the algorithms in the NISQ era but also for achieving higher classification accuracy.

\end{document}